\documentclass[11pt]{article}

\usepackage[letterpaper,margin=1in]{geometry}
\usepackage[utf8]{inputenc}
\usepackage[T1]{fontenc}
\usepackage{lmodern}
\usepackage[numbers,sort&compress]{natbib}
\usepackage{amsmath,amssymb,amsfonts,amsthm,mathtools}
\usepackage{booktabs}
\usepackage{array}
\usepackage{multirow}

\usepackage{graphicx}
\usepackage{subcaption}
\usepackage{float}
\graphicspath{{figures/}}
\usepackage{microtype}
\usepackage{hyperref}
\usepackage{url}

\hypersetup{
  hidelinks,
  pdftitle={Predicting Only from Selected Evidence: A Tempered Product-of-Experts Bottleneck for Auditable EEG Diagnosis},
  pdfauthor={Yinghao WANG, Shujian YU, Duc Han LE, Zhikai YU, Changming WANG, Van-Tam NGUYEN}
}
\title{Predicting Only from Selected Evidence: A Tempered Product-of-Experts Bottleneck for Auditable EEG Diagnosis}
\author{Yinghao WANG\textsuperscript{1}, Shujian YU\textsuperscript{2,*}, Duc Han LE\textsuperscript{1}, Zhikai YU\textsuperscript{3}, Changming WANG\textsuperscript{3,*}, Van-Tam NGUYEN\textsuperscript{1}}
\newcommand{\authoraffiliations}{%
  \textsuperscript{1}LTCI, T\'el\'ecom Paris, Institut Polytechnique de Paris, Palaiseau, France\\
  \textsuperscript{2}Department of Artificial Intelligence, Vrije Universiteit Amsterdam, Amsterdam, The Netherlands\\
  \textsuperscript{3}Beijing Preclinical Testing Platform for Brain--Computer Interface Products,\\
  Zhongguancun Industrial Park, Beijing, China\\[0.35em]
  \textsuperscript{*}\emph{Co-corresponding authors.}
}
\date{}

\makeatletter
\renewcommand{\maketitle}{%
  \begin{center}
    {\Large\bfseries \@title\par}
    \vspace{0.9em}
    {\small\itshape \@author\par}
    \vspace{0.65em}
    {\footnotesize \authoraffiliations\par}
  \end{center}
  \vspace{1.25em}
}
\makeatother

\newcommand{\R}{\mathbb{R}}
\newcommand{\E}{\mathbb{E}}
\newcommand{\KL}{D_{\mathrm{KL}}}
\newcommand{\N}{\mathcal{N}}

\newcommand{\diag}{\mathrm{diag}}

\newcommand{\ce}{\mathcal{L}_{\mathrm{cls}}}
\newcommand{\lkl}{\mathcal{L}_{\mathrm{KL}}}
\newcommand{\lsparse}{\mathcal{L}_{\mathrm{sparse}}}
\newcommand{\lcons}{\mathcal{L}_{\mathrm{cons}}}

\theoremstyle{plain}
\newtheorem{proposition}{Proposition}
\newtheorem{corollary}{Corollary}
\newtheorem{lemma}{Lemma}

\theoremstyle{remark}
\newtheorem{remark}{Remark}

\begin{document}
\maketitle

\begin{abstract}
Pretrained EEG backbones improve transfer performance, but downstream diagnosis heads remain hard to audit: predictions are made from unrestricted hidden states, whereas explanations are usually produced only after the decision. We introduce tPoE-EIB, an \emph{evidence-information bottleneck} head for adapting EEG backbones under an evidence-only prediction constraint. tPoE-EIB selects temporal and channel evidence, maps the selected summaries to Gaussian experts over a shared latent variable, and fuses them with a tempered product-of-experts posterior. The classifier observes only this latent, so the decision path is explicit and rate-limited by the expected posterior KL. This gives a tractable supervised objective with an information-rate penalty, while the closed-form tempered posterior mitigates overconfident fusion from correlated evidence axes. We evaluate tPoE-EIB on pretrained EEG foundation-model backbones across six diagnosis settings: event-type classification, abnormality detection, seizure detection, cognitive-decline staging, depression screening, and cerebrovascular-disease classification. The evaluation spans public benchmarks and in-house clinical cohorts, binary screening and fine-grained staging, and sparse and dense montages. tPoE-EIB preserves competitive balanced accuracy and improves over representative post-hoc explanations on selection-faithfulness audits, including insertion--deletion and gate-causality tests. Its structured posterior further enables integration-faithfulness audits, including expert-drop, posterior-reliance, and expert-disagreement tests. Overall, these results suggest that evidence-only, rate-limited fusion is a practical route to auditable diagnosis on top of frozen EEG foundation backbones.
\end{abstract}

\section{Introduction}
Deep EEG modeling has shifted from compact task-specific networks toward pretrained foundation-style backbones. Architectures such as EEGNet and EEG Conformer introduced EEG-specific convolutional and attention biases \citep{lawhern2018eegnet,song2023eegconformer}, while recent models such as LaBraM, CBraMod, and CodeBrain learn reusable EEG representations through large-scale pretraining and structured temporal--channel modeling \citep{jiang2024labram,wang2024cbramod,ma2025codebrain}. As representation quality improves, however, the downstream diagnosis head becomes the main bottleneck for auditability. We use \emph{auditability} in a narrow, operational sense: the evidence path should be explicit enough to intervene on and measure. A frozen backbone followed by pooling and a task head may perform well, but the classifier can still exploit any predictive regularity in the hidden state, including nuisance structure, subject-specific signatures, or artifact-correlated cues. Explanations are then produced only after the prediction has already been fixed.

Post-hoc explainers such as Integrated Gradients, Dynamask, and TIMING identify salient inputs after a prediction \citep{sundararajan2017ig,crabbe2021dynamask,jang2025timing}, but they do not constrain the information used to make that prediction. This separation between explanation generation and model optimization raises a faithfulness problem: a saliency map need not reflect the computation that actually drove the decision \citep{jacovi2020towards}. In high-stakes settings, this concern has motivated calls for models whose interpretability is built into the decision path rather than attached afterward \citep{rudin2019stop}. The issue is especially salient in EEG diagnosis, where clinicians inspect evidence along two natural axes: when a relevant event occurs and where on the scalp it appears. A model that spreads support diffusely across time, or concentrates support on anatomically implausible channels, may be accurate but difficult to audit.

Selection alone is not enough. A sparse or stable gate is not automatically faithful: a gate that assigns the same low-mass pattern to every input can satisfy sparsity and augmentation consistency while selecting no sample-specific evidence. Thus, in addition to selecting temporal and channel support, a downstream head must restrict what the selected evidence can carry and expose the resulting path to intervention-based audits \citep{yu2019rethinking}.

\begin{figure}[t]
\centering
\includegraphics[width=\linewidth]{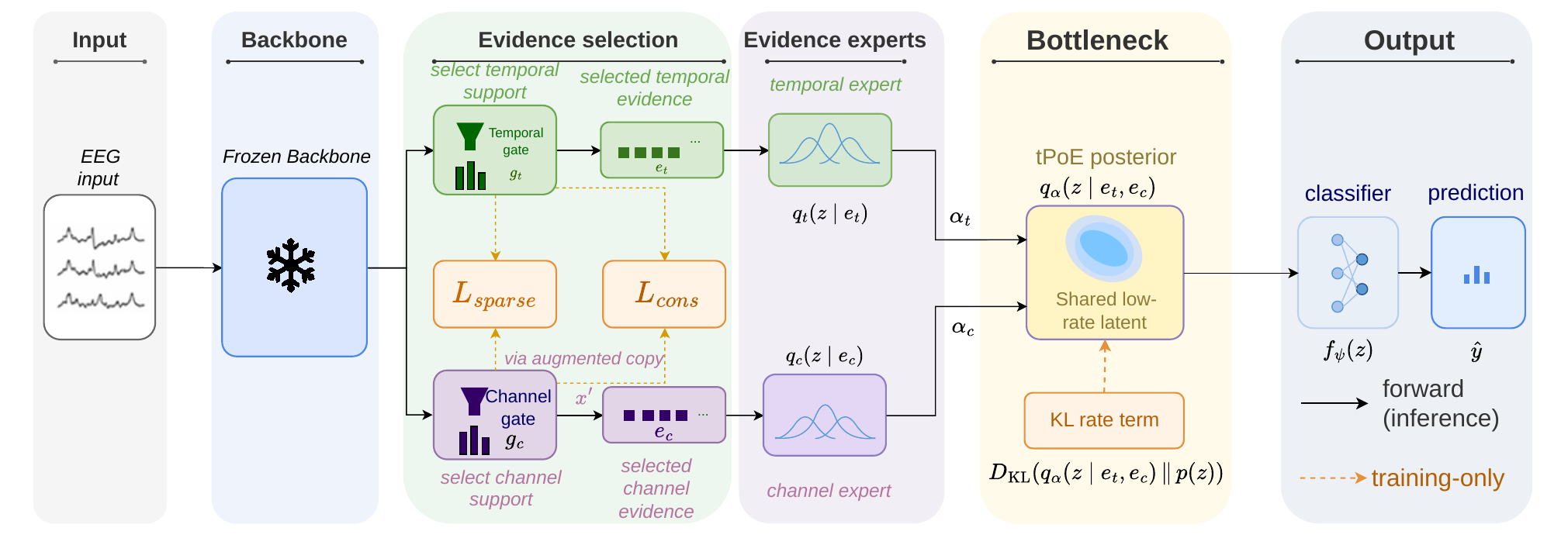}
\caption{Architecture of tPoE-EIB. Solid arrows denote the inference path; 
orange dashed arrows denote training-only regularization. A frozen EEG 
backbone produces a patch-grid representation $H$. Temporal and channel 
gates $g_t, g_c$ select evidence summaries $e_t$ and $e_c$, which two 
Gaussian experts map to beliefs $q_t(z\mid e_t)$ and $q_c(z\mid e_c)$ 
over the same latent variable. These are integrated by a tempered 
product-of-experts posterior $q_\alpha(z\mid e_t,e_c) \propto p(z)\,
q_t^{\alpha_t} q_c^{\alpha_c}$ with axis-specific temperatures 
$\alpha_t,\alpha_c\in[0,1]$, and the classifier $f_\psi$ observes only 
the latent $z$. Two selection-stage regularizers shape the gates: 
$\mathcal{L}_{\mathrm{sparse}}$ penalizes gate mass to prevent trivial 
all-pass solutions, and $\mathcal{L}_{\mathrm{cons}}$ encourages stable 
gates by comparing $g_t(x), g_c(x)$ to the gates obtained on a lightly 
augmented copy $x'$ (augmented branch omitted from the diagram for 
clarity). By construction, the classifier never receives the 
concatenated evidence $[e_t;e_c]$ or the unrestricted hidden state $H$;
all classification information flows through the rate-limited latent 
$z$.}
\label{fig:architecture}
\end{figure}

We propose \textbf{tPoE-EIB}, a downstream adaptation head that puts evidence on the forward decision path (Fig.~\ref{fig:architecture}). Given a frozen backbone representation, tPoE-EIB learns temporal and channel gates, pools the gated support into axis-specific evidence summaries, and maps those summaries to Gaussian experts over a shared latent variable. A tempered product-of-experts posterior integrates the experts, and the classifier receives only a sample from this latent during training and its posterior mean at evaluation. We call this restriction an \emph{evidence-only prediction constraint}: after the frozen backbone, the classifier cannot access pooled hidden states or concatenated evidence summaries, and any feature that affects the prediction must pass through the gates, the evidence summaries, and the tPoE posterior. The latent is rate-limited because the KL term controls how much selected evidence can reach the classifier. Clinical correctness is not claimed by construction and is evaluated separately through the audits in Sec.~\ref{sec:eval}.

We evaluate tPoE-EIB on CodeBrain backbone, with cross-backbone validation on CBraMod, across EEG diagnosis settings that include event-type classification, abnormality detection, seizure detection, cognitive decline staging, depression, and cerebrovascular disease. The tasks span public benchmarks and in-house clinical cohorts, binary screening regimes and a fine-grained staging setting, and both sparse and dense channel montages.

Our contributions are as follows:
\begin{itemize}
    \item \textbf{Evidence-only EEG adaptation.} We formulate downstream adaptation of pretrained EEG backbones as an \emph{evidence-only prediction constraint}: the diagnosis head cannot access unrestricted pooled backbone features and must predict from gated, rate-limited evidence.

    \item \textbf{Tempered PoE evidence bottleneck.} We instantiate this constraint with a tempered product-of-experts bottleneck in which temporal and channel evidence summaries become Gaussian experts over a shared latent variable and are integrated by precision-weighted posterior aggregation rather than feature concatenation.

    \item \textbf{Theory and audit protocol.} We derive the closed-form tPoE posterior, connect the training loss to a variational lower bound on an evidence-IB objective, prove a rate bound on the stochastic decision path, and use a Gaussian calibration model to motivate tempering for correlated evidence axes. We further introduce audits for gate faithfulness, expert necessity, gate causality, posterior reliance, and temporal--channel conflict.

    \item \textbf{Empirical validation across EEG diagnosis settings.} Across public benchmarks and in-house clinical datasets, tPoE-EIB remains competitive with unrestricted heads while exposing decision paths that can be audited at both the selection and posterior-integration stages.
\end{itemize}
\section{Related Work}
\paragraph{Pretrained EEG backbones and adaptation heads.}
EEG-specific networks such as EEGNet and EEG Conformer introduced compact convolutional and attention biases for multichannel EEG decoding \citep{lawhern2018eegnet,song2023eegconformer}. Recent foundation-style EEG models shift the emphasis to reusable representations: LaBraM learns generic EEG representations through large-scale masked prediction, CBraMod models temporal--channel structure under heterogeneous EEG formats, CodeBrain decouples time--frequency tokenization from multi-scale representation learning, and concurrent variants explore decoder-centric seq2seq pretraining and topology-aware mixture-of-experts designs \citep{jiang2024labram,wang2024cbramod,ma2025codebrain,liu2025echo,chen2025uni}. These backbones improve transfer, but common downstream heads---linear probes, MLP heads, and attention pooling---still expose unrestricted hidden states to the classifier. We focus on this last step: how a frozen EEG representation is allowed to enter the diagnosis head.
\paragraph{EEG interpretability and intrinsic constraints.}
EEG interpretability often relies on attention visualization or post-hoc saliency. Integrated Gradients, Dynamask, and TIMING can highlight samples or time regions after prediction, but the classifier may still have used information outside the displayed explanation \citep{sundararajan2017ig,crabbe2021dynamask,jang2025timing,jacovi2020towards}. Intrinsic approaches constrain the prediction path itself: Right for the Right Reasons regularizes explanation gradients, Concept Bottleneck Models route predictions through concepts, and rationale methods train selectors for sufficient input subsets \citep{ross2017right,koh2020cbm,zarlenga2022cem,yuksekgonul2023posthoc,oikarinen2023label,lei2016rationalizing,bastings2019rationale}. tPoE-EIB is closer to intrinsic bottlenecking than to post-hoc attribution, but its intermediate variables are not human-labeled concepts. They are model-selected temporal and channel evidence summaries, chosen along axes that EEG reviewers already inspect. Because selectors can collapse to task-independent templates \citep{yu2019rethinking}, we audit both the gates and the posterior that consumes their selected evidence.
\paragraph{Information bottlenecks and posterior fusion.}
The information bottleneck principle compresses representations while preserving label-relevant information, and Deep VIB makes this trade-off tractable with a variational KL rate term \citep{tishby2000ib,alemi2017vib,shwartz2017opening}. Multi-view and multimodal variants study how correlated sources should be retained or fused \citep{federici2020learning,zhu2020mmved,song2021mmib,han2022dynamics,jiamultimodal}. Simple concatenation does not reveal which source is informative for a given sample. Product-of-experts fusion gives a posterior aggregation view, but standard PoE treats experts as conditionally independent. In our setting, temporal and channel summaries are correlated projections of the same EEG representation, so we temper the PoE to discount duplicated support rather than treating the two axes as independent modalities.
\section{tPoE-EIB}
\label{sec:method}

\subsection{Problem setup and evidence-only prediction constraint}
\label{subsec:problem_setup}
Let $x\in\R^{C\times S\times P}$ denote an EEG input with $C$ channels, $S$ temporal segments, and $P$ samples per segment. A frozen pretrained backbone $B_\theta$ maps $x$ to a patch-grid representation $H=B_\theta(x)\in\R^{C\times S\times D}$. Standard adaptation pools $H$ and trains a classifier on the resulting unrestricted representation. tPoE-EIB instead inserts an evidence interface between backbone and classifier:
\begin{equation}
H=B_\theta(x),\qquad (g_t,g_c,z)=E_\phi(H),\qquad \hat y=f_\psi(z),
\label{eq:forward}
\end{equation}
where $g_t\in[0,1]^S$, $g_c\in[0,1]^C$ are gates and $z\in\R^{d_z}$ is a stochastic decision latent. The classifier receives only $z$: neither the gates nor the evidence summaries $e_t,e_c$ defined below are exposed to $f_\psi$. We use \emph{evidence-only prediction constraint} to denote this forward-graph restriction. It gives forward-path faithfulness to the model-selected evidence: if a feature affects $\hat y$, it must pass through the gates, the evidence summaries, and the posterior bottleneck. It does not assert that the selected evidence is clinically correct, so Sec.~\ref{sec:eval} evaluates the path with intervention-based audits.

\subsection{Evidence-information bottleneck objective and training surrogate}
\label{subsec:eib_objective}
Let $E=(e_t,e_c)$ collect the temporal and channel evidence summaries produced from $H$ (Sec.~\ref{subsec:dual_axis}), and let $q_\alpha(z\mid E)$ be the stochastic encoder that maps this evidence to $z$ (Sec.~\ref{subsec:tpoe}). We optimize the \emph{evidence-information bottleneck} objective
\begin{equation}
\mathcal{J}_{\mathrm{EIB}}=I(Y;Z)-\beta\,I(E;Z),
\label{eq:eib}
\end{equation}
which replaces the raw input in the original IB \citep{tishby2000ib,alemi2017vib} by the selected evidence $E$.
\paragraph{Why compress $E$ rather than $H$?}
The classifier never receives the full hidden state $H$; it receives $Z$ sampled from $q_\alpha(z\mid E(H))$. The implemented graph is therefore $H\to E\to Z\to\hat Y$. Under this graph, compressing $I(E;Z)$ controls the information on the only path available to the classifier. A direct penalty on $I(H;Z)$ would be a coarser hidden-state rate term: under the deterministic evidence map it coincides with this path-wise quantity, and more generally it upper-bounds it by data processing, but it would obscure the fact that the head is designed to regulate selected evidence rather than the entire frozen representation.

\begin{proposition}[Variational lower bound on $\mathcal{J}_{\mathrm{EIB}}$]\label{prop:eib}
For any decoder $r_\psi(y\mid z)$ and prior $p(z)=\N(0,I)$,
\begin{align}
\mathcal{J}_{\mathrm{EIB}}\;\ge\;
\E_{p(x,y)}\E_{q_\alpha(z\mid E)}[\log r_\psi(y\mid z)]
-\beta\,\E_{p(x)}\KL(q_\alpha(z\mid E)\Vert p(z))
+H(Y).
\label{eq:eib_bound}
\end{align}
\end{proposition}

The proof is in Appendix~\ref{app:eib_proof}. Maximizing the bound is equivalent to minimizing $\ce=-\E[\log r_\psi(y\mid z)]$ plus a KL rate term $\lkl=\E_{p(x)}\KL(q_\alpha(z\mid E)\Vert p(z))$. The full training objective adds two selection-stage regularizers acting on the gates,
\begin{equation}
\mathcal{L}_{\mathrm{total}}=\ce+\beta_{\mathrm{eff}}\,\lkl+\eta\,\lsparse+\gamma\,\lcons,
\label{eq:loss}
\end{equation}
with $\lsparse,\lcons$ defined in Eqs.~\eqref{eq:sparse}--\eqref{eq:cons} and $\beta_{\mathrm{eff}}$ linearly warmed up over the first $T_{\mathrm{warm}}$ epochs. The KL term has an operational interpretation: by the data-processing inequality on $E\to Z\to\hat Y$, $I(\hat Y;X)\le \E_{p(E)}\KL(q_\alpha(z\mid E)\Vert p(z))$, so the trained KL upper-bounds how much input information can affect the prediction along the stochastic training path (Corollary~\ref{cor:rate} in Appendix~\ref{app:theory}).
\subsection{Dual-axis evidence selection}
\label{subsec:dual_axis}
From $H$, tPoE-EIB forms marginal temporal and channel views by averaging over the complementary axis,
$H_t[s] = \tfrac{1}{C}\sum_{c}H_{c,s,:}$ and $H_c[j] = \tfrac{1}{S}\sum_{s}H_{j,s,:}$, so $H_t\in\R^{S\times D}$ and $H_c\in\R^{C\times D}$. Two small MLPs predict continuous sigmoid gates $g_t=\sigma(\mathrm{MLP}_t(H_t))$ and $g_c=\sigma(\mathrm{MLP}_c(H_c))$, broadcast over feature dimensions, giving evidence summaries
\begin{equation}
e_t=\frac{1}{S}\sum_{s=1}^{S}g_t[s]H_t[s],\qquad
e_c=\frac{1}{C}\sum_{j=1}^{C}g_c[j]H_c[j].
\label{eq:evidence}
\end{equation}
The factorization is intentionally transparent: marginal temporal and channel support rather than a full time--channel evidence map. A concatenation head would feed $[e_t;e_c]$ directly to $f_\psi$; tPoE-EIB removes this path. The selection-stage regularizers act on these gates,
\begin{align}
\lsparse&=\E_{p(x)}\!\left[\tfrac{1}{2}\!\left(\tfrac{1}{S}\textstyle\sum_{s} g_t(x)[s]+\tfrac{1}{C}\sum_{j} g_c(x)[j]\right)\right],\label{eq:sparse}\\
\lcons&=\E_{p(x)}\!\left[\tfrac{1}{2}\!\left(\tfrac{1}{S}\|g_t(x)-g_t(x')\|_2^2+\tfrac{1}{C}\|g_c(x)-g_c(x')\|_2^2\right)\right],\label{eq:cons}
\end{align}
where $x'=\mathrm{Aug}(x)$ is a light augmentation. $\lsparse$ limits how much support enters the evidence interface; the tPoE KL in Eq.~\eqref{eq:loss} limits how much selected evidence reaches the classifier. Lemma~\ref{lem:constant} (Appendix~\ref{app:theory}) shows that $\lsparse$ and $\lcons$ alone do not guarantee sample-dependent selection, motivating the rate-limited bottleneck and the audits in Sec.~\ref{sec:eval}.
\subsection{Tempered Product-of-Experts posterior}
\label{subsec:tpoe}
Temporal and channel summaries are treated as Gaussian evidence experts over the same latent $z$:
\begin{equation}
q_t(z\mid e_t)=\N(\mu_t(e_t),\diag(\sigma_t^2(e_t))),\qquad
q_c(z\mid e_c)=\N(\mu_c(e_c),\diag(\sigma_c^2(e_c))).
\label{eq:experts}
\end{equation}
Both experts have dimension $d_z$ and share the prior $p(z)=\N(0,I)$. Because temporal and channel summaries are correlated projections of the same EEG representation, standard PoE can treat duplicated support as independent evidence and become overconfident; Proposition~\ref{prop:overconf} below formalizes this in a Gaussian calibration model. We therefore use a \emph{tempered} product:
\begin{equation}
q_\alpha(z\mid e_t,e_c)\propto p(z)\,q_t(z\mid e_t)^{\alpha_t}q_c(z\mid e_c)^{\alpha_c},
\qquad \alpha_t,\alpha_c\in[0,1].
\label{eq:tpoe}
\end{equation}
Setting $\alpha_t=\alpha_c=1$ recovers standard PoE; we also evaluate a learnable variant constrained to $[0,1]$ (Appendix~\ref{sec:impl}).

\begin{proposition}[Tempering corrects overconfidence under correlated experts]\label{prop:overconf}
Let $z\in\R$ have prior $p(z)=\N(0,1)$, and let two experts observe noisy views $y_t = z + \varepsilon_t$ and $y_c = z + \varepsilon_c$, where $(\varepsilon_t,\varepsilon_c)$ are jointly Gaussian with zero mean, marginal variances $\sigma^2$, and correlation $\rho\in[0,1)$. Treating the experts as Gaussian \emph{likelihood factors}, $q_t(z;y_t)\propto \exp\!\big(-(z-y_t)^2/(2\sigma^2)\big)$ and $q_c(z;y_c)\propto \exp\!\big(-(z-y_c)^2/(2\sigma^2)\big)$, the true Bayesian posterior $p(z\mid y_t,y_c)\propto p(z)\,p(y_t,y_c\mid z)$ has precision $\tau^*=1+2/(\sigma^2(1+\rho))$. The tempered PoE with $\alpha_t=\alpha_c=\alpha$ has posterior precision $\tau_\alpha=1+2\alpha/\sigma^2$, and the unique $\alpha^*$ matching the true posterior is
\begin{equation}
\alpha^*=\frac{1}{1+\rho}\in\left(\tfrac{1}{2},1\right].
\label{eq:alpha_star}
\end{equation}
Thus standard PoE ($\alpha=1$) is calibrated only when $\rho=0$ and is overconfident whenever $\rho>0$; for any $\rho>0$, a tempered configuration $\alpha^*\in(1/2,1)$ matches the true posterior precision.
\end{proposition}

\begin{remark}
This is a calibration argument for posterior precision, not a claim that the learned neural experts are exact Bayesian likelihoods.
\end{remark}

The proof is in Appendix~\ref{app:overconf_proof}. Proposition~\ref{prop:overconf} also explains the conservative fixed choice $\alpha=0.5$: this is the limiting calibrated temperature as $\rho\to 1$, the worst case for duplicated evidence. Since the actual correlation is unknown and sample-dependent, we treat $0.5$ as a lower endpoint rather than a universally optimal temperature. The aggregated posterior is Gaussian in closed form.

\begin{proposition}[Closed-form tempered PoE posterior]\label{prop:tpoe}
Under the experts in Eq.~\eqref{eq:experts} with diagonal covariance and $p(z)=\N(0,I)$, $q_\alpha(z\mid e_t,e_c)=\N(\mu_\alpha,\diag(\sigma_\alpha^2))$ with
\begin{align}
\tau_{\alpha,j}&=1+\alpha_t\sigma_{t,j}^{-2}+\alpha_c\sigma_{c,j}^{-2},\quad \sigma_{\alpha,j}^{2}=\tau_{\alpha,j}^{-1},\label{eq:tau}\\
\mu_{\alpha,j}&=\sigma_{\alpha,j}^{2}\left(\alpha_t\sigma_{t,j}^{-2}\mu_{t,j}+\alpha_c\sigma_{c,j}^{-2}\mu_{c,j}\right).\label{eq:tpoemean}
\end{align}
\end{proposition}

The proof is in Appendix~\ref{app:tpoe_proof}. The posterior mean is a precision-weighted convex combination of the prior mean $0$ and the two expert means with weights
\begin{equation}
w_{t,j}=\frac{\alpha_t\sigma_{t,j}^{-2}}{\tau_{\alpha,j}},\quad
w_{c,j}=\frac{\alpha_c\sigma_{c,j}^{-2}}{\tau_{\alpha,j}},\quad
w_{0,j}=\frac{1}{\tau_{\alpha,j}},\quad w_{t,j}+w_{c,j}+w_{0,j}=1.
\label{eq:reliance_weights}
\end{equation}
An axis contributes strongly only when its expert is confident and its temperature is not suppressed; these weights define the posterior-reliance audit in Sec.~\ref{sec:eval}. During training we sample
\begin{equation}
z=\mu_\alpha+\sigma_\alpha\odot\epsilon,\qquad \epsilon\sim\N(0,I),
\label{eq:sample}
\end{equation}
and at evaluation $z=\mu_\alpha$. The classifier $f_\psi$ is a small MLP on this latent.
\section{Evaluation Protocol}
\label{sec:eval}
We use \emph{audit} in an intervention sense: each metric perturbs one part of the evidence path and measures the effect on the trained model. The five metrics target distinct questions and are not interchangeable.

\textbf{(i) Gate faithfulness.} For axis $a\in\{t,c\}$, segments or channels are ranked by descending gate value on the clean input; insertion progressively restores high-ranked units to a fully masked input and deletion progressively zeros them in the original. The faithfulness gap is $\mathrm{gap}_a=\mathrm{AUC}^{(a)}_{\mathrm{ins}}-\mathrm{AUC}^{(a)}_{\mathrm{del}}$; a positive gap means the gate ranking is useful under perturbation, not that the selected units are clinically sufficient.

\textbf{(ii) Expert necessity.} Setting $\alpha_c=0$ collapses the channel expert into the prior ($q_\alpha\propto p(z)\,q_t^{\alpha_t}$); $\alpha_t=0$ is the channel-only analogue. We report $\Delta_a=M_{\mathrm{full}}-M_{\alpha_a=0}$ for a predictive metric $M$.

\textbf{(iii) Gate causality.} For each axis and sweep value $v\in[0,1]$, we replace $g_a(x)$ by the constant $v\mathbf{1}$ at evaluation time and read off $M(v)$. A flat curve indicates an irrelevant gate value; a cliff indicates a causally necessary one. Unlike expert-drop, gate-causality overrides the gate while keeping the trained expert head in place.

\textbf{(iv) Posterior reliance.} The precision weights from Eq.~\eqref{eq:reliance_weights} give $\rho_a=\tfrac{1}{d_z}\sum_j w_{a,j}$, partitioning the posterior into temporal, channel, and prior contributions: gates say \emph{where} evidence was selected; reliance says \emph{how much} posterior precision each axis contributes after selection.

\textbf{(v) Expert disagreement.} $D_{\mathrm{SKL}}(q_t,q_c)=\tfrac{1}{2}[\KL(q_t\Vert q_c)+\KL(q_c\Vert q_t)]$ is large when the two axes support different beliefs about $z$; we treat it as an expert-conflict diagnostic, not an uncertainty score.

Detailed protocols (perturbation grid, masking, diagonal-Gaussian form of $D_{\mathrm{SKL}}$) are in Appendix~\ref{app:eval_protocol}.
\section{Experiments}
\label{sec:experiments}
We organize the experiments around four questions. Q1 (Sec.~\ref{subsec:main_results}) asks whether the evidence-only constraint remains predictively feasible across backbones: does denying the head access to pooled hidden states cost accuracy, and does the answer transfer from CodeBrain to CBraMod? Q2 (Sec.~\ref{subsec:ablations}) ablates the choice of head, the temperature, the KL bottleneck, and the gate-shaping regularizers. Q3 (Sec.~\ref{subsec:audit_results}) turns to the audits---insertion--deletion, expert-drop, posterior reliance, expert disagreement, and gate-causality sweeps---and asks what the restricted forward graph makes visible. Q4 (Sec.~\ref{subsec:clinical_alignment}) tests whether the selected channels recover clinically motivated regional anchors.
\subsection{Experimental setup}
\label{subsec:setup}
\paragraph{Setup.}
We evaluate on six EEG diagnosis tasks: event-type classification (TUEV), abnormality detection (TUAB), seizure detection (TUSZ)\citep{obeid2016theg}, cognitive decline staging (AD/MCI/SCD/HC), depression, and cerebrovascular disease. The in-house cohorts use $58$--$129$-channel EGI HydroCel and BrainVision grids (Appendix~\ref{app:datasets}), denser than the 16--22-channel TUH public benchmarks. All splits are subject-disjoint before windowing. Our primary frozen backbone is CodeBrain \citep{ma2025codebrain}; CBraMod \citep{wang2024cbramod} is a cross-backbone check on the subset of tasks with matched preprocessing and completed 3-seed runs. We compare four heads that differ only in temporal/channel integration: \emph{Linear probe}, \emph{Axis-Concat}, \emph{PoE-EIB} ($\alpha_t{=}\alpha_c{=}1$, standard PoE bottleneck), and \emph{tPoE-EIB} (learnable $\alpha_t,\alpha_c\in[0,1]$). Unless otherwise stated, $d_z{=}64$, $\beta{=}10^{-4}$ with five-epoch warmup, $\eta{=}10^{-3}$, $\gamma{=}10^{-2}$, 50 epochs of AdamW (lr $10^{-3}$), patience 15, and 3 seeds. Hyperparameter and preprocessing details are in Appendices~\ref{app:datasets}--\ref{sec:impl}.
\subsection{Q1: Performance under the evidence-only constraint}
\label{subsec:main_results}
Table~\ref{tab:external_baselines} reports balanced accuracy and matched temporal/channel faithfulness gaps $\mathrm{gap}_t,\mathrm{gap}_c$ for tPoE-EIB and a panel of CNN, self-supervised, and pretrained EEG baselines. The comparison is not a broad leaderboard; it asks two coupled questions: (i)~does denying the head access to pooled hidden states cost accuracy, and (ii)~does the answer transfer when we swap the frozen backbone?
\paragraph{Linear vs.\ tPoE-EIB across backbones.}
Replacing the linear probe with tPoE-EIB improves balanced accuracy on every public-benchmark cell, with gains of $+1.7/+4.3$ pts on TUAB, $+2.1/+2.7$ on TUEV, and $+2.2/+2.7$ on TUSZ (CodeBrain / CBraMod). The absolute gain is consistently larger on the weaker CBraMod backbone, suggesting the evidence-only interface is most valuable when the underlying representation is weaker.
\paragraph{Comparison to external methods.}
CodeBrain-tPoE-EIB attains the best BAcc on the three public benchmarks: TUAB ($80.7\%$), TUEV ($48.8\%$), and TUSZ ($76.4\%$). On the in-house cohorts, pretrained foundation backbones do not dominate: ContraWR leads on Depression ($65.7\%$) and Cerebrovascular ($64.2\%$), and BIOT ties with CodeBrain-tPoE-EIB on AD staging ($38.5\%$). tPoE-EIB on either backbone exceeds the corresponding linear probe on every clinical cohort, so the evidence-only constraint does not sacrifice accuracy even when other architectures lead in absolute terms.

\begin{table}[t]
\centering
\caption{Cross-model comparison across six EEG diagnosis tasks, split into two row-blocks (public benchmarks above, in-house clinical cohorts below). Each task reports 3-seed balanced accuracy (\%) and matched temporal/channel faithfulness gaps $\mathrm{gap}_t,\mathrm{gap}_c$(integrated gradients on the posterior, uniform across methods). Best BAcc per task is in \textbf{bold}.}
\label{tab:external_baselines}
\scriptsize
\setlength{\tabcolsep}{3pt}
\begin{tabular}{@{}l*{9}{c}@{}}
\toprule
& \multicolumn{3}{c}{TUAB} & \multicolumn{3}{c}{TUEV} & \multicolumn{3}{c}{TUSZ} \\
\cmidrule(lr){2-4} \cmidrule(lr){5-7} \cmidrule(lr){8-10}
Method & BAcc & $\mathrm{gap}_t$ & $\mathrm{gap}_c$ & BAcc & $\mathrm{gap}_t$ & $\mathrm{gap}_c$ & BAcc & $\mathrm{gap}_t$ & $\mathrm{gap}_c$ \\
\midrule
EEGNet              & $79.4{\pm}0.6$         & $0.129$ & $-0.005$ & $27.8{\pm}2.1$         & $0.020$ & $-0.012$ & $50.2{\pm}0.0$         & $0.001$ & $0.001$  \\
ContraWR            & $78.3{\pm}1.5$         & $0.159$ & $0.140$  & $47.9{\pm}2.7$ & $0.151$ & $0.066$  & $61.7{\pm}3.8$         & $0.014$ & $0.068$  \\
BENDR               & $63.3{\pm}1.3$         & $0.055$ & $0.064$  & $16.7{\pm}0.0$         & $-0.026$ & $0.003$ & $50.0{\pm}0.0$         & $0.000$ & $0.000$  \\
BIOT                & $71.6{\pm}0.1$         & $0.074$ & $0.003$  & $37.4{\pm}0.1$         & $0.087$ & $-0.055$ & $60.9{\pm}6.2$         & $0.015$ & $0.002$  \\
LaBraM              & $73.2{\pm}0.1$         & $0.176$ & $-0.019$ & $40.6{\pm}0.3$         & $0.092$ & $-0.028$ & $55.4{\pm}0.6$         & $0.017$ & $-0.006$ \\
CodeBrain-Linear    & $79.0{\pm}0.3$         & $0.058$ & $0.085$  & $46.7{\pm}0.6$         & $0.150$ & $-0.052$ & $74.2{\pm}0.6$         & $0.127$ & $-0.025$ \\
CBraMod-Linear      & $68.2{\pm}0.6$         & $0.042$ & $-0.255$ & $30.4{\pm}2.4$         & $0.093$ & $-0.087$ & $57.6{\pm}1.2$         & $0.027$     & $-0.008$     \\
CodeBrain-tPoE-EIB  & $\mathbf{80.7{\pm}0.2}$ & $0.169$ & $0.066$  & $\mathbf{48.8{\pm}0.7}$         & $0.152$ & $-0.003$ & $\mathbf{76.4{\pm}0.5}$ & $0.098$ & $-0.010$ \\
CBraMod-tPoE-EIB    & $72.5{\pm}0.2$         & $0.020$ & $-0.119$ & $33.1{\pm}1.7$         & $0.118$ & $-0.097$ & $60.3{\pm}1.7$         & $0.031$     & $-0.006$     \\
\bottomrule
\end{tabular}

\vspace{4pt}

\begin{tabular}{@{}l*{9}{c}@{}}
\toprule
& \multicolumn{3}{c}{AD staging} & \multicolumn{3}{c}{Depression} & \multicolumn{3}{c}{Cerebrovascular} \\
\cmidrule(lr){2-4} \cmidrule(lr){5-7} \cmidrule(lr){8-10}
Method & BAcc & $\mathrm{gap}_t$ & $\mathrm{gap}_c$ & BAcc & $\mathrm{gap}_t$ & $\mathrm{gap}_c$ & BAcc & $\mathrm{gap}_t$ & $\mathrm{gap}_c$ \\
\midrule
EEGNet              & $22.6{\pm}6.4$         & $0.030$  & $0.011$  & $63.0{\pm}4.2$         & $0.024$  & $0.011$  & $56.0{\pm}0.4$         & $-0.004$ & $-0.029$ \\
ContraWR            & $23.6{\pm}3.3$         & $0.180$  & $-0.058$ & $\mathbf{65.7{\pm}1.7}$ & $0.215$  & $0.046$  & $\mathbf{64.2{\pm}1.2}$ & $-0.035$ & $-0.049$ \\
BENDR               & $25.4{\pm}0.4$         & $-0.002$ & $0.001$  & $56.3{\pm}1.0$         & $0.110$  & $0.013$  & $50.1{\pm}0.1$         & $-0.004$ & $0.001$  \\
BIOT                & $\mathbf{38.5{\pm}2.0}$ & $0.036$  & $0.008$  & $59.2{\pm}0.8$         & $0.173$  & $0.254$  & $61.4{\pm}0.2$         & $-0.034$ & $-0.007$ \\
LaBraM              & $36.0{\pm}1.3$         & $0.074$  & $-0.018$ & $61.2{\pm}0.1$         & $0.184$  & $0.042$  & $54.8{\pm}0.4$         & $0.033$  & $0.027$  \\
CodeBrain-Linear    & $31.0{\pm}1.6$         & $0.222$  & $-0.078$ & $54.2{\pm}0.9$         & $0.276$  & $-0.091$ & $56.8{\pm}0.8$         & $0.184$  & $-0.008$ \\
CBraMod-Linear      & $25.2{\pm}0.2$         & $0.215$  & $-0.032$ & $50.5{\pm}0.4$         & $0.085$  & $0.059$  & $54.8{\pm}0.8$         & $0.294$  & $-0.088$ \\
CodeBrain-tPoE-EIB  & $\mathbf{38.5{\pm}4.1}$         & $0.340$  & $-0.072$ & $58.8{\pm}2.0$         & $0.221$  & $0.365$ & $58.0{\pm}0.8$         & $0.290$  & $-0.047$ \\
CBraMod-tPoE-EIB    & $26.1{\pm}5.4$         & $0.023$  & $0.001$  & $54.3{\pm}0.4$         & $-0.033$ & $0.052$  & $57.1{\pm}1.4$         & $-0.028$ & $0.014$  \\
\bottomrule
\end{tabular}
\end{table}
\subsection{Q2: Ablations on heads, tempering, and regularizers}
\label{subsec:ablations}
\paragraph{Head ablation.}
Table~\ref{tab:main} fills in the four-head sweep on a frozen CodeBrain backbone, completing the Linear-vs.-tPoE-EIB picture from Q1 with the two intermediate variants Axis-Concat and PoE-EIB ($\alpha=1$). tPoE-EIB ranks first on five of seven tasks (TUAB, TUEV, TUSZ, AD staging, Cerebrovascular) and second on AD binary and Depression, where standard PoE without tempering is best. Among the three non-Linear heads, the per-task spread is at most $2.0$ pts on the public benchmarks and at most $7.2$ pts on the in-house cohorts; gate selection and rate-limited posterior therefore contribute on the same scale rather than one dominating. Calibration is task-dependent (Appendix~\ref{app:calibration_full}): tPoE-EIB attains the lowest NLL on all three public benchmarks (TUAB $0.427$, TUEV $0.882$, TUSZ $0.227$) and the lowest ECE on TUEV ($0.117$) and TUSZ ($0.018$); only TUAB ECE favors the unrestricted Axis-Concat head ($0.018$ vs.\ $0.029$). The bottleneck therefore does not impose a consistent calibration cost---and on TUEV it cuts NLL by $0.338$ relative to Axis-Concat.

\begin{table}[t]
\centering
\caption{Head ablation: balanced accuracy (\%, mean $\pm$ std over 3 seeds) on EEG diagnosis tasks with a frozen CodeBrain backbone. Best per task is in \textbf{bold}; second best is \underline{underlined}.}
\label{tab:main}
\small
\begin{tabular}{@{}lcccc@{}}
\toprule
Dataset & Linear probe & Axis-Concat & PoE-EIB ($\alpha=1$) & tPoE-EIB (learnable $\alpha$) \\
\midrule
TUAB & $79.0{\pm}0.3$ & $\underline{80.4{\pm}0.5}$ & $80.3{\pm}0.4$ & $\mathbf{80.7{\pm}0.2}$ \\
TUEV & ${46.7{\pm}0.6}$ & $46.8{\pm}1.4$ & $\underline{47.1{\pm}0.5}$ & $\mathbf{48.8{\pm}0.7}$ \\
TUSZ & $74.2{\pm}0.6$ & $75.1{\pm}0.9$ & $\underline{75.9{\pm}0.6}$ & $\mathbf{76.4{\pm}0.5}$ \\
AD staging & $28.5{\pm}4.1$ & $31.3{\pm}3.1$ & $\underline{34.0{\pm}1.6}$ & $\mathbf{38.5{\pm}4.1}$ \\
AD binary & $75.2{\pm}1.6$ & $75.6{\pm}3.6$ & $\mathbf{77.5{\pm}2.6}$ & $\underline{77.4{\pm}4.8}$ \\
Depression & $54.2{\pm}0.9$ & ${58.2{\pm}2.0}$ & $\mathbf{60.4{\pm}1.1}$ & $\underline{58.8{\pm}2.0}$ \\
Cerebrovascular & $56.8{\pm}0.8$ & $57.4{\pm}0.8$ & $\underline{57.6{\pm}1.0}$ & $\mathbf{58.0{\pm}0.8}$ \\
\bottomrule
\end{tabular}
\end{table}
\paragraph{Temperature ablation.}
Comparing fixed $\alpha\in\{0.5,1.0\}$ to a learnable $\alpha\in[0,1]$ via sigmoid reparameterization (Appendix~\ref{app:temperature}), the learnable variant attains the highest balanced accuracy on all three public benchmarks and converges to $\alpha\approx 0.5$ on both axes, near the worst-case calibrated temperature $\alpha^*=1/(1+\rho)$ derived in Proposition~\ref{prop:overconf} for highly correlated experts ($\rho\to 1$).
\paragraph{Loss and gate-shaping ablations.}
The pattern is channel-count-dependent (Appendix~\ref{app:loss_ablation}): on TUAB, single-regularizer ablations leave $\mathrm{gap}_c\in[+0.051,+0.061]$ and BAcc within $0.5$ pts; on 129-channel Depression, removing any regularizer flips $\mathrm{gap}_c$ from $+0.104$ to between $-0.113$ and $-0.190$, so channel-axis compression is essential at high channel count. On TUAB, the Axis-Concat control's temporal gate range is $0.137$ vs.\ $0.316$ for tPoE-EIB---the gate stays a light pass-through without the posterior bottleneck (Appendix~\ref{app:gate_nuisance}).
\subsection{Q3: Auditing selected evidence and posterior integration}
\label{subsec:audit_results}
We test whether the restricted forward graph produces evidence signals that can be audited along five axes: gate faithfulness, expert necessity, posterior reliance, expert disagreement, and gate causality. These audits answer different questions, so we avoid reducing them to a single explanation score.
\paragraph{Gate faithfulness (insertion--deletion).}
\begin{figure}[t]
\centering
\includegraphics[width=0.7\linewidth]{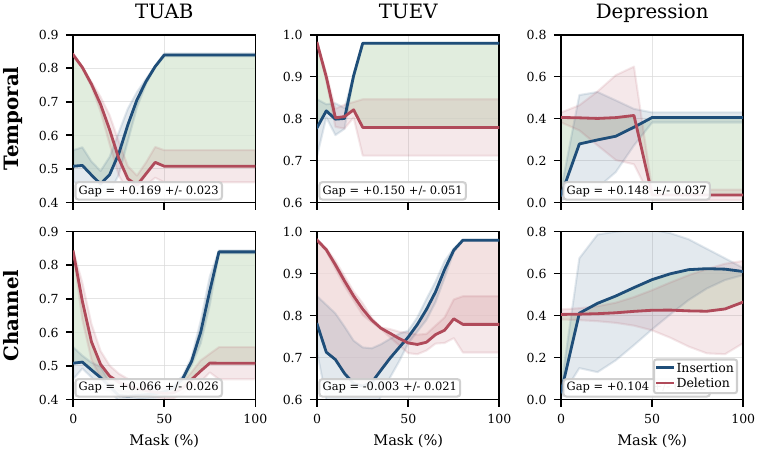}
\caption{Insertion--deletion curves on TUAB, TUEV, and Depression. The shaded area between insertion (top) and deletion (bottom) curves quantifies gate faithfulness on each axis.}
\label{fig:audit_visuals}
\end{figure}
TUAB shows positive gaps on both axes (temporal $0.169{\pm}0.023$, channel $0.066{\pm}0.026$). On TUEV, the temporal gap remains positive ($0.150{\pm}0.051$) while the channel gap is near zero ($-0.003{\pm}0.021$), consistent with pathological-event discrimination depending on localized temporal structure rather than a stable per-channel ranking.
\paragraph{Expert necessity, reliance, disagreement.}
Table~\ref{tab:audit} summarizes the audits on three representative tasks; gaps here are computed from the learned tPoE-EIB gates, while Table~\ref{tab:external_baselines} uses integrated gradients on the posterior (uniform across all methods). Expert-drop identifies channel as the dominant axis on TUAB ($\Delta_c=30.1$ vs.\ $\Delta_t=22.6$ pts), and the learnable $\alpha$ uses both experts more evenly than standard PoE. Posterior reliance shows $\rho_0\approx0.52$--$0.62$, confirming the bottleneck rate-limits evidence flow; reliance correlates with $\mathrm{gap}_c$ but not $\mathrm{gap}_t$, so we read it as a posterior-use diagnostic rather than a substitute for perturbation faithfulness. Expert disagreement $D_{\mathrm{SKL}}$ runs counter to the usual ``high uncertainty $\to$ wrong'' reading: on TUAB, correctly classified samples carry \emph{higher} $D_{\mathrm{SKL}}$, so disagreement should be read as temporal--channel conflict rather than confidence (Appendix~\ref{app:reliance_stratified}).

A complementary gate-causality sweep that distinguishes axis regimes between CVD and Depression is presented in Appendix~\ref{app:gate_causality_regime}.

\begin{table}[t]
\centering
\caption{Audit summary on representative tasks (best-validation checkpoint of tPoE-EIB). }
\label{tab:audit}
\small
\begin{tabular}{@{}lcccccccc@{}}
\toprule
Dataset & $\mathrm{gap}_t$ & $\mathrm{gap}_c$ & $\Delta_t$ & $\Delta_c$ & $\rho_t$ & $\rho_c$ & $\rho_0$ & $D_{\mathrm{SKL}}$ \\
\midrule
TUAB & $+0.161$ & $+0.061$ & $0.226$ & $0.301$ & $0.14$ & $0.24$ & $0.62$ & $445\mathrm{k}$ \\
TUEV & $+0.150$ & $-0.003$ & $0.145$ & $0.224$ & $0.18$ & $0.23$ & $0.59$ & $68k$ \\
TUSZ & $+0.098$ & $-0.010$ & $0.130$ & $0.269$ & $0.18$ & $0.30$ & $0.52$ & $42\mathrm{k}$ \\
\bottomrule
\end{tabular}
\end{table}
\subsection{Q4: Clinical alignment of selected evidence}
\label{subsec:clinical_alignment}
\begin{figure}[t]
\centering
\begin{minipage}[b]{0.32\linewidth}
\centering
\includegraphics[width=\linewidth]{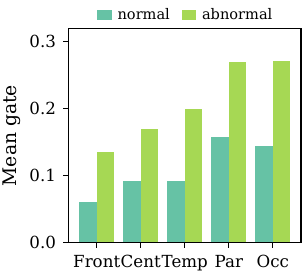}
\subcaption{TUAB: correct $\times$ class}
\label{fig:tuab_align}
\end{minipage}\hfill
\begin{minipage}[b]{0.32\linewidth}
\centering
\includegraphics[width=\linewidth]{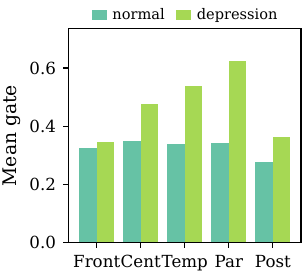}
\subcaption{Depression: correct $\times$ class}
\label{fig:dep_align}
\end{minipage}\hfill
\begin{minipage}[b]{0.32\linewidth}
\centering
\includegraphics[width=\linewidth]{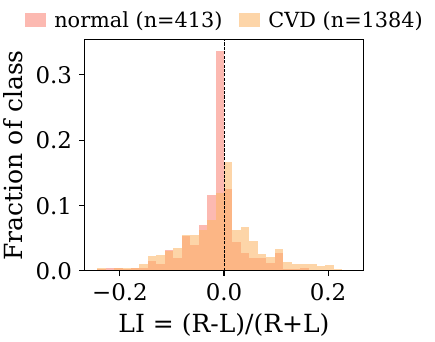}
\subcaption{Cerebrovascular: $\mathrm{LI}$}
\label{fig:cvd_align}
\end{minipage}
\caption{Clinical alignment visualizations on three tasks (correctly classified samples only, split by ground-truth class). (a) TUAB: both classes peak at parietal/occipital channels consistent with posterior dominant rhythm, but the abnormal class recruits roughly $2{\times}$ the gate magnitude of the normal class at every region, with the largest absolute gap at occipital and parietal sites. (b) Depression: the depressive class shows substantially elevated gate at parietal ($0.62$ vs.\ $0.34$), central, and temporal sites, while frontal weight is similar between classes---the discriminator is parietal/central elevation rather than the classical frontal alpha asymmetry alone. (c) Cerebrovascular disease: CVD samples show a substantially broader lateralization-index distribution than normal controls, consistent with focal lesion-side slowing. Per-task region distributions, the full \emph{correct $\times$ class} grid, and per-class profiles are in Appendix~\ref{app:clinical_alignment_figs}.}
\label{fig:clinical_alignment_main}
\end{figure}

The Q3 audits test whether the gate is mechanistically faithful, but not whether selected channels recover the \emph{task-specific anatomical biomarker} established by clinical EEG. For each task we pre-register a regional hypothesis from the textbook EEG signature, average the per-sample channel gate within canonical 10--20 region groups (frontal, central, temporal, parietal, posterior/occipital), and run a one-sided Wilcoxon test of the predicted target--control contrast (or of $|\mathrm{LI}|=|(L-R)/(L+R)|$ for focal-slowing tasks). Tests are repeated over three seeds. Table~\ref{tab:clinical_alignment} reports outcomes and Fig.~\ref{fig:clinical_alignment_main} visualizes three representative tasks using the cleaner \emph{correct-only, class-conditioned} view (samples that the model classified correctly, then split by ground-truth class). The pattern is task-specific and reproducible across seeds.

\paragraph{Hypothesis recovered (TUAB, Depression, CVD).}
On TUAB, the selector concentrates on parieto-occipital sites with the abnormal class consistently doubling the normal-class gate magnitude, in line with posterior-dominant-rhythm disruption \citep{lodder2011pdr,marcuse2008pdr} ($d{=}1.75{\pm}0.37$, $p{<}10^{-3}$). Depression elevates central/parietal/temporal weight in the depressive class with frontal exceeding posterior in the directional Wilcoxon contrast ($d{=}2.06{\pm}1.36$); the parietal involvement extends beyond the classical FAA frame \citep{henriques1991left,allen2015faa,vanderVinne2017faa}. CVD shows pronounced lateralization and a posterior shift specific to the disease class, consistent with focal lesion-side slowing \citep{finnigan2013strokeqeeg,finnigan2016dar} ($|\mathrm{LI}|$ effect $d{=}4.92{\pm}2.67$, $p{<}10^{-3}$).
\paragraph{TUEV per-event profile.}
TUEV does not admit a single directional contrast, so we report the per-class profile. Among correctly predicted events, EYEM peaks at frontal channels ($0.78$ vs.\ $0.46$--$0.74$ at other regions), consistent with the Fp1/Fp2/F7/F8 origin of ocular activity. Periodic-discharge classes carry the highest gate overall: GPED concentrates at parietal/central sites, while PLED is broad with a frontal peak. ARTF carries the lowest weight at every region, consistent with down-weighting non-physiological windows (Fig.~\ref{fig:correct_by_class_grid}f).
\paragraph{Hypothesis rejected, accuracy preserved (TUSZ, AD).}
TUSZ reaches $76.4{\pm}0.5\%$ but the focal-onset hypothesis (temporal/frontal $>$ central/parietal) is rejected in every seed ($d{=}-0.41{\pm}0.19$); the seizure-class profile (Fig.~\ref{fig:correct_by_class_grid}d) recruits gate broadly with a central/occipital peak, so the bottleneck found a more general solution than the lesion-localized prior anticipated for the heterogeneous TUSZ event mix. AD attains best-in-class balanced accuracy among heads ($38.5{\pm}4.1\%$, chance $25\%$; tied with BIOT at $38.5\%$ in Table~\ref{tab:external_baselines}), yet the posterior--frontal contrast goes the wrong direction in every seed ($d{=}-6.54{\pm}4.46$) and the per-region gate saturates near $1.0$ at 4-way granularity (Fig.~\ref{fig:region_gate_means_grid}d). A binary AD vs.\ NC reformulation recovers the gate's discriminative content---a post-hoc logistic regression on the 58-dim channel gate reaches $88.8$--$97.9\%$ balanced accuracy with a pronounced fronto-parietal gate-correlation reversal (Appendix~\ref{app:ad_nc_binary})---so the 4-way failure reflects task granularity, not absent AD-relevant signal; the recovered pattern is distributed rather than aligned with the pre-registered ``posterior $>$ frontal'' anchor \citep{jeong2004eegad,lizio2011eegad}.

Alignment outcome and predictive accuracy therefore decouple, and an unrestricted Axis-Concat head cannot draw these distinctions because its classifier never commits to a region.

\begin{table}[t]
\centering
\caption{Clinical alignment of channel selectors against task-specific EEG biomarkers. Cohen's $d$ is the directional target--control effect size (or the $|\mathrm{LI}|$ effect for CVD), reported as 3-seed mean $\pm$ std on the held-out test split. \emph{Aligned} summarizes the one-sided Wilcoxon decision across seeds; TUEV uses a per-event-class analysis instead of a directional contrast. Test-set sample counts in Table~\ref{tab:datasets}.}
\label{tab:clinical_alignment}
\small
\setlength{\tabcolsep}{4pt}
\begin{tabular}{@{}llcc@{}}
\toprule
Task & Anchor / predicted gate pattern & $d$ (3-seed) & Aligned \\
\midrule
TUAB        & PDR: par./occ.\ $>$ front.                                            & $+1.75{\pm}0.37$ & \checkmark\ (3/3) \\
TUSZ        & Focal onset: temp./front.\ $>$ cent./par.                              & $-0.41{\pm}0.19$ & $\times$ (3/3) \\
TUEV        & Per-event (EYEM\,$\to$\,front.; GPED/PLED\,$\to$\,cent.)               & per-class       & EYEM front.\ ($0.78$) \\
AD staging  & Post.\ $\alpha\!\downarrow$/$\theta\delta\!\uparrow$: post.\ $>$ front. & $-6.54{\pm}4.46$ & $\times$ (3/3) \\
Depression  & FAA: front.\ $>$ post.                                                 & $+2.06{\pm}1.36$ & \checkmark\ (3/3) \\
Cerebrovascular & Lesion-side slowing: $|\mathrm{LI}|\!\uparrow$ in disease class    & $+4.92{\pm}2.67$ & \checkmark\ (3/3) \\
\bottomrule
\end{tabular}
\end{table}
\section{Conclusion}
\label{sec:conclusion}
We introduced tPoE-EIB, an evidence-information bottleneck head that adapts pretrained EEG backbones under an evidence-only prediction constraint, with a variational lower bound on an evidence-IB criterion and a tempered product of Gaussian temporal and channel experts over a shared low-rate latent. The interface is intentionally factorized---marginal temporal and channel support rather than arbitrary joint patterns---so it cannot directly express evidence such as ``channel $j$ matters only during segment $s$,'' and group-level audits have blind spots when gate values are nearly uniform or highly patient-specific. Three directions preserving the evidence-only restriction extend this framework: rate-adaptive bottlenecks that allocate $d_z$ by class or task, hierarchical experts that decompose multi-class decisions into nested judgments, and joint time--channel masks with structured priors that capture patterns marginal gates cannot express.

\newpage

\bibliographystyle{unsrtnat}
\bibliography{references}

@inproceedings{alemi2017vib,
  author    = {Alemi, Alexander A. and Fischer, Ian and Dillon, Joshua V. and Murphy, Kevin},
  title     = {Deep variational information bottleneck},
  booktitle = {International Conference on Learning Representations},
  year      = {2017}
}

@inproceedings{bastings2019rationale,
  author    = {Bastings, Jasmijn and Aziz, Wilker and Titov, Ivan},
  title     = {Interpretable neural predictions with differentiable binary variables},
  booktitle = {Proceedings of the 57th Annual Meeting of the Association for Computational Linguistics},
  year      = {2019}
}

@inproceedings{crabbe2021dynamask,
  author    = {Crabb{\'e}, Jonathan and van der Schaar, Mihaela},
  title     = {Explaining time series predictions with dynamic masks},
  booktitle = {Proceedings of the 38th International Conference on Machine Learning},
  year      = {2021}
}

@inproceedings{federici2020learning,
  author    = {Federici, Marco and Dutta, Anjan and Forr{\'e}, Patrick and Kushman, Nate and Akata, Zeynep},
  title     = {Learning robust representations via multi-view information bottleneck},
  booktitle = {International Conference on Learning Representations},
  year      = {2020}
}

@inproceedings{han2022dynamics,
  author    = {Han, Zongbo and Yang, Fan and Huang, Junzhou and Zhang, Changqing and Yao, Jianhua},
  title     = {Multimodal dynamics: Dynamical fusion for trustworthy multimodal classification},
  booktitle = {Proceedings of the IEEE/CVF Conference on Computer Vision and Pattern Recognition},
  year      = {2022}
}

@inproceedings{jang2025timing,
  author    = {Jang, Hyeongwon and Kim, Changhun and Yang, Eunho},
  title     = {{TIMING}: Temporality-aware integrated gradients for time series explanation},
  booktitle = {Proceedings of the 42nd International Conference on Machine Learning},
  year      = {2025}
}

@inproceedings{jiang2024labram,
  author    = {Jiang, Wei-Bang and Zhao, Li-Ming and Lu, Bao-Liang},
  title     = {Large brain model for learning generic representations with tremendous {EEG} data in {BCI}},
  booktitle = {International Conference on Learning Representations},
  year      = {2024}
}

@inproceedings{koh2020cbm,
  author    = {Koh, Pang Wei and Nguyen, Thao and Tang, Yew Siang and Mussmann, Stephen and Pierson, Emma and Kim, Been and Liang, Percy},
  title     = {Concept bottleneck models},
  booktitle = {Proceedings of the 37th International Conference on Machine Learning},
  year      = {2020}
}

@article{lawhern2018eegnet,
  author    = {Lawhern, Vernon J. and Solon, Amelia J. and Waytowich, Nicholas R. and Gordon, Stephen M. and Hung, Chou P. and Lance, Brent J.},
  title     = {{EEGNet}: A compact convolutional neural network for {EEG}-based brain--computer interfaces},
  journal   = {Journal of Neural Engineering},
  volume    = {15},
  number    = {5},
  pages     = {056013},
  year      = {2018}
}

@inproceedings{lei2016rationalizing,
  author    = {Lei, Tao and Barzilay, Regina and Jaakkola, Tommi},
  title     = {Rationalizing neural predictions},
  booktitle = {Proceedings of the 2016 Conference on Empirical Methods in Natural Language Processing},
  year      = {2016}
}

@inproceedings{ma2025codebrain,
  title={Codebrain: Bridging decoupled tokenizer and multi-scale architecture for eeg foundation model},
  author={Ma, Jingying and Wu, Feng and Lin, Qika and Xing, Yucheng and Liu, Chenyu and Jia, Ziyu and Feng, Mengling},
  booktitle={The Fourteenth International Conference on Learning Representations},
  year={2025}
}

@inproceedings{oikarinen2023label,
  author    = {Oikarinen, Tuomas and Das, Subhro and Nguyen, Lam M. and Weng, Tsui-Wei},
  title     = {Label-free concept bottleneck models},
  booktitle = {International Conference on Learning Representations},
  year      = {2023}
}

@inproceedings{ross2017right,
  author    = {Ross, Andrew Slavin and Hughes, Michael C. and Doshi-Velez, Finale},
  title     = {Right for the right reasons: Training differentiable models by constraining their explanations},
  booktitle = {Proceedings of the Twenty-Sixth International Joint Conference on Artificial Intelligence},
  year      = {2017}
}

@article{shwartz2017opening,
  author    = {Shwartz-Ziv, Ravid and Tishby, Naftali},
  title     = {Opening the black box of deep neural networks via information},
  journal   = {arXiv preprint arXiv:1703.00810},
  year      = {2017}
}

@article{song2021mmib,
  author    = {Song, Jingqi and Zheng, Yuanjie and Wang, Jing and Ullah, Muhammad Zakir and Jiao, Wanzhen},
  title     = {Multicolor image classification using the multimodal information bottleneck network ({MMIB}-{N}et) for detecting diabetic retinopathy},
  journal   = {Optics Express},
  volume    = {29},
  number    = {14},
  pages     = {22732--22748},
  year      = {2021}
}

@article{song2023eegconformer,
  author    = {Song, Yunfa and Zheng, Qian and Liu, Bin and Gao, Xiaorong},
  title     = {{EEG} Conformer: Convolutional transformer for {EEG} decoding and visualization},
  journal   = {IEEE Transactions on Neural Systems and Rehabilitation Engineering},
  volume    = {31},
  pages     = {710--719},
  year      = {2023}
}

@inproceedings{sundararajan2017ig,
  author    = {Sundararajan, Mukund and Taly, Ankur and Yan, Qiqi},
  title     = {Axiomatic attribution for deep networks},
  booktitle = {Proceedings of the 34th International Conference on Machine Learning},
  year      = {2017}
}

@inproceedings{tishby2000ib,
  author    = {Tishby, Naftali and Pereira, Fernando C. and Bialek, William},
  title     = {The information bottleneck method},
  booktitle = {Proceedings of the 37th Annual Allerton Conference on Communication, Control and Computing},
  pages     = {368--377},
  year      = {2000}
}

@inproceedings{wang2024cbramod,
    title={{CB}raMod: A Criss-Cross Brain Foundation Model for {EEG} Decoding},
    author={Jiquan Wang and Sha Zhao and Zhiling Luo and Yangxuan Zhou and Haiteng Jiang and Shijian Li and Tao Li and Gang Pan},
    booktitle={The Thirteenth International Conference on Learning Representations},
    year={2025}
}

@inproceedings{yu2019rethinking,
  author    = {Yu, Mo and Chang, Shiyu and Zhang, Yang and Jaakkola, Tommi},
  title     = {Rethinking cooperative rationalization: Introspective extraction and complement control},
  booktitle = {Proceedings of the 2019 Conference on Empirical Methods in Natural Language Processing and the 9th International Joint Conference on Natural Language Processing},
  year      = {2019}
}

@inproceedings{yuksekgonul2023posthoc,
  author    = {Yuksekgonul, Mert and Wang, Maggie and Zou, James},
  title     = {Post-hoc concept bottleneck models},
  booktitle = {International Conference on Learning Representations},
  year      = {2023}
}

@inproceedings{zarlenga2022cem,
  author    = {Espinosa Zarlenga, Mateo and Barbiero, Pietro and Ciravegna, Gabriele and Marra, Giuseppe and Giannini, Francesco and Diligenti, Michelangelo and Shams, Zohreh and Precioso, Frederic and Melacci, Stefano and Weller, Adrian and Lio, Pietro and Jamnik, Mateja},
  title     = {Concept embedding models: Beyond the accuracy-explainability trade-off},
  booktitle = {Advances in Neural Information Processing Systems},
  year      = {2022}
}

@article{zhu2020mmved,
  author    = {Zhu, Yaochen and Xie, Jiayi and Chen, Zhenzhong},
  title     = {Predicting the popularity of micro-videos with multimodal variational encoder-decoder framework},
  journal   = {arXiv preprint arXiv:2003.12724},
  year      = {2020}
}

@article{rudin2019stop,
     title={Stop explaining black box machine learning models for high stakes decisions and use interpretable models instead},
     author={Rudin, Cynthia},
     journal={Nature Machine Intelligence},
     volume={1}, number={5}, pages={206--215}, year={2019}
   }

@inproceedings{jacovi2020towards,
     title={Towards Faithfully Interpretable {NLP} Systems: How Should We Define and Evaluate Faithfulness?},
     author={Jacovi, Alon and Goldberg, Yoav},
     booktitle={Proceedings of the 58th Annual Meeting of the Association for Computational Linguistics (ACL)},
     pages={4198--4205}, year={2020}
   }

@article{obeid2016theg,
  title={The temple university hospital EEG data corpus},
  author={Obeid, Iyad and Picone, Joseph},
  journal={Frontiers in neuroscience},
  volume={10},
  pages={196},
  year={2016},
  publisher={Frontiers Media SA}
}

@article{lodder2011pdr,
  author    = {Lodder, Shaun S. and van Putten, Michel J. A. M.},
  title     = {Automated {EEG} analysis: characterizing the posterior dominant rhythm},
  journal   = {Journal of Neuroscience Methods},
  volume    = {200},
  number    = {1},
  pages     = {86--93},
  year      = {2011}
}

@article{marcuse2008pdr,
  author    = {Marcuse, Lara V. and Schneider, Marc and Mortati, Katherine A. and Donnelly, Karen M. and Arnedo, Vanessa and Grant, Arthur C.},
  title     = {Quantitative analysis of the {EEG} posterior-dominant rhythm in healthy adolescents},
  journal   = {Clinical Neurophysiology},
  volume    = {119},
  number    = {8},
  pages     = {1778--1781},
  year      = {2008}
}

@article{jeong2004eegad,
  author    = {Jeong, Jaeseung},
  title     = {{EEG} dynamics in patients with {A}lzheimer's disease},
  journal   = {Clinical Neurophysiology},
  volume    = {115},
  number    = {7},
  pages     = {1490--1505},
  year      = {2004}
}

@article{lizio2011eegad,
  author    = {Lizio, Roberta and Vecchio, Fabrizio and Frisoni, Giovanni B. and Ferri, Raffaele and Rodriguez, Guido and Babiloni, Claudio},
  title     = {Electroencephalographic rhythms in {A}lzheimer's disease},
  journal   = {International Journal of Alzheimer's Disease},
  volume    = {2011},
  pages     = {927573},
  year      = {2011}
}

@article{henriques1991left,
  author    = {Henriques, Jeffrey B. and Davidson, Richard J.},
  title     = {Left frontal hypoactivation in depression},
  journal   = {Journal of Abnormal Psychology},
  volume    = {100},
  number    = {4},
  pages     = {535--545},
  year      = {1991}
}

@article{allen2015faa,
  author    = {Allen, John J. B. and Reznik, Samantha J.},
  title     = {Frontal {EEG} asymmetry as a promising marker of depression vulnerability: summary and methodological considerations},
  journal   = {Current Opinion in Psychology},
  volume    = {4},
  pages     = {93--97},
  year      = {2015}
}

@article{vanderVinne2017faa,
  author    = {van der Vinne, Nikita and Vollebregt, Madelon A. and van Putten, Michel J. A. M. and Arns, Martijn},
  title     = {Frontal alpha asymmetry as a diagnostic marker in depression: fact or fiction? {A} meta-analysis},
  journal   = {NeuroImage: Clinical},
  volume    = {16},
  pages     = {79--87},
  year      = {2017}
}

@article{finnigan2013strokeqeeg,
  author    = {Finnigan, Simon and van Putten, Michel J. A. M.},
  title     = {{EEG} in ischaemic stroke: quantitative {EEG} can uniquely inform (sub-)acute prognoses and clinical management},
  journal   = {Clinical Neurophysiology},
  volume    = {124},
  number    = {1},
  pages     = {10--19},
  year      = {2013}
}

@article{finnigan2016dar,
  author    = {Finnigan, Simon and Wong, Andrew and Read, Stephen},
  title     = {Defining abnormal slow {EEG} activity in acute ischaemic stroke: delta/alpha ratio as an optimal {QEEG} index},
  journal   = {Clinical Neurophysiology},
  volume    = {127},
  number    = {2},
  pages     = {1452--1459},
  year      = {2016}
}

@article{liu2025echo,
  title={ECHO: Toward Contextual Seq2Seq Paradigms in Large EEG Models},
  author={Liu, Chenyu and Deng, Yuqiu and Liu, Tianyu and Zhou, Jinan and Zhou, Xinliang and Jia, Ziyu and Ding, Yi},
  journal={arXiv preprint arXiv:2509.22556},
  year={2025}
}

@article{chen2025uni,
  title={Uni-{NTFM}: A Unified Foundation Model for {EEG} Signal Representation Learning},
  author={Chen, Zhisheng and Zhang, Yingwei and Lan, Qizhen and Liu, Tianyu and Wang, Huacan and Ding, Yi and Jia, Ziyu and Chen, Ronghao and Wang, Kun and Zhou, Xinliang},
  journal={arXiv preprint arXiv:2509.24222},
  year={2025}
}

@inproceedings{jiamultimodal,
  title={A Multimodal {BiMamba} Network with Test-Time Adaptation for Emotion Recognition Based on Physiological Signals},
  author={Jia, Ziyu and Du, Tingyu and Tian, Zhengyu and Li, Hongkai and Zhang, Yong and Liu, Chenyu},
  booktitle={Advances in Neural Information Processing Systems},
  year={2025}
}
\newpage
\appendix

\section{Conceptual overview}
\label{app:teaser}

\begin{figure}[H]
\centering
\includegraphics[width=\linewidth]{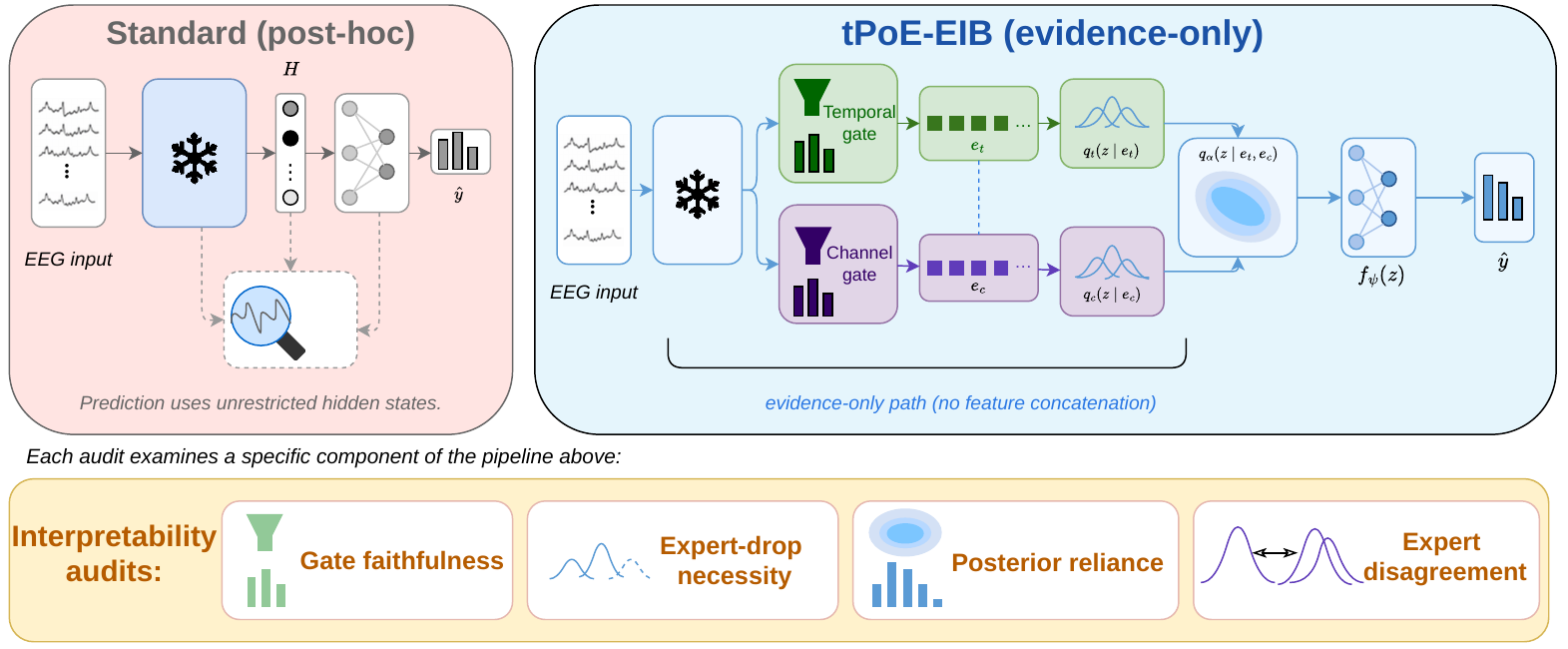}
\caption{Conceptual contrast between a standard post-hoc downstream head and tPoE-EIB. A standard head (left) predicts from unrestricted hidden states $H$ and explains the decision afterward. tPoE-EIB (right) routes prediction through temporal and channel gates, evidence summaries $e_t,e_c$, Gaussian experts $q_t(z\mid e_t)$ and $q_c(z\mid e_c)$, and a tempered product-of-experts posterior $q_\alpha(z\mid e_t,e_c)$. The classifier $f_\psi$ observes only the latent, so no concatenated evidence or unrestricted hidden state reaches the classifier.}
\label{fig:teaser}
\end{figure}

For reference, Fig.~\ref{fig:teaser} contrasts a standard post-hoc downstream head with the evidence-only forward path used by tPoE-EIB.

\section{Theoretical properties (extended)}
\label{app:theory}

This appendix expands the theoretical statements referenced in the main text. We record two further properties: a rate-limited path bound that follows from the variational bound in Proposition~\ref{prop:eib}, and a constant-gate failure mode (Lemma~\ref{lem:constant}) that motivates the audits in Sec.~\ref{sec:eval}. We also state a variance-conservatism corollary that complements the calibration argument for tempering given by Proposition~\ref{prop:overconf} in the main text.

\subsection{Rate-limited evidence path}

\begin{corollary}[Rate-limited evidence path]\label{cor:rate}
Along the forward graph $E\to Z\to \hat Y$, the data-processing inequality gives
\begin{equation*}
I(\hat Y;E)\le I(Z;E)\le \E_{p(E)}\KL\!\big(q_\alpha(z\mid E)\,\Vert\,p(z)\big).
\end{equation*}
Because $X\to H\to E$ is deterministic under the frozen backbone and deterministic gates, the same stochastic path bounds the dependence of the prediction on the input: $I(\hat Y;X)\le \E_{p(E)}\KL(q_\alpha(z\mid E)\Vert p(z))$.
\end{corollary}

\begin{remark}
This information-theoretic statement applies to the stochastic training path $Z\sim q_\alpha(z\mid E)$ in Eq.~\eqref{eq:sample}. At evaluation, we use the posterior mean $\mu_\alpha$ as a variance-reduced predictor. This preserves the architectural restriction that the classifier sees only the tPoE latent parameterization, but the mutual-information bound itself is tied to the stochastic training path.
\end{remark}

\subsection{Stability is not selectivity}
Sparsity and consistency penalties on the gates do not by themselves ensure sample-dependent evidence selection: a constant gate can minimize both.

\begin{lemma}[Stability is not selectivity]\label{lem:constant}
Let $g_t(x)=\bar g_t$ and $g_c(x)=\bar g_c$ be constant for all $x$. Then the consistency loss in Eq.~\eqref{eq:cons} is zero for any augmentation operator $\mathrm{Aug}$. The sparsity loss in Eq.~\eqref{eq:sparse} equals $\tfrac{1}{2}\bigl(\|\bar g_t\|_1/S+\|\bar g_c\|_1/C\bigr)$, which depends only on the chosen constants and is small whenever those constants are sparse.
\end{lemma}

The proof is in Appendix~\ref{app:lemma_proof}. Lemma~\ref{lem:constant} formalizes the constant-gate failure: such a gate is a degenerate rationale in the sense of \citet{yu2019rethinking}. The bottleneck does not make this shortcut impossible, but Corollary~\ref{cor:rate} ensures that any information passing through a collapsed selector is still rate-limited, and Sec.~\ref{sec:eval} turns that constraint into measurable evidence-quality tests.

\subsection{Variance conservatism of tempered PoE}
Proposition~\ref{prop:overconf} (in the main text) gives the calibration argument for tempering. A complementary observation is that tempering is also conservative with respect to posterior variance.

\begin{corollary}[Tempering is conservative relative to standard PoE]\label{cor:conservative}
If $0\le \alpha_t,\alpha_c\le 1$, then the tPoE posterior variance is element-wise no smaller than the standard PoE variance obtained with $\alpha_t=\alpha_c=1$. This follows directly from Eqs.~\eqref{eq:tau}.
\end{corollary}

\section{Additional Proofs}
\label{app:additional_proofs}
\subsection{Proof of Lemma~\ref{lem:constant} (stability is not selectivity)}
\label{app:lemma_proof}
For every $x$, $g_t(\mathrm{Aug}(x))=\bar g_t=g_t(x)$ and similarly for $g_c$, so the squared differences in Eq.~\eqref{eq:cons} vanish. The sparsity claim is immediate from Eq.~\eqref{eq:sparse} since both averages equal $\|\bar g_t\|_1/S$ and $\|\bar g_c\|_1/C$.\qed

\subsection{Proof of Proposition~\ref{prop:tpoe} (closed-form tempered PoE posterior)}
\label{app:tpoe_proof}
Let $\Sigma_t=\diag(\sigma_t^2)$ and $\Sigma_c=\diag(\sigma_c^2)$ denote the diagonal covariance matrices of the two experts. Expanding $\log q_\alpha(z)$ from Eq.~\eqref{eq:tpoe} gives the sum of three quadratic forms in $z$: the standard normal prior term $-\tfrac{1}{2}\|z\|^2$ and the two tempered expert log-densities $-\tfrac{\alpha_t}{2}(z-\mu_t)^\top\Sigma_t^{-1}(z-\mu_t)$ and $-\tfrac{\alpha_c}{2}(z-\mu_c)^\top\Sigma_c^{-1}(z-\mu_c)$. Collecting quadratic and linear terms in $z$ yields a Gaussian with precision matrix $I+\alpha_t\Sigma_t^{-1}+\alpha_c\Sigma_c^{-1}$ and natural parameter $\alpha_t\Sigma_t^{-1}\mu_t+\alpha_c\Sigma_c^{-1}\mu_c$. Diagonal covariance gives Eqs.~\eqref{eq:tau}--\eqref{eq:tpoemean}.\qed

\subsection{Proof of Proposition~\ref{prop:eib} (variational lower bound)}
\label{app:eib_proof}
We bound the two terms of $\mathcal{J}_{\mathrm{EIB}}=I(Y;Z)-\beta I(E;Z)$ separately.

\emph{Label term (lower bound on $I(Y;Z)$).}
By definition, $I(Y;Z)=H(Y)-H(Y\mid Z)=H(Y)+\E_{p(y,z)}[\log p(y\mid z)]$. For any decoder $r_\psi$,
$\E_{p(y,z)}[\log p(y\mid z)]=\E_{p(y,z)}[\log r_\psi(y\mid z)]+\E_{p(z)}\KL(p(y\mid z)\Vert r_\psi(y\mid z))$,
and the KL term is non-negative. Therefore
\begin{equation}
I(Y;Z)\ge H(Y)+\E_{p(x,y)}\E_{q_\alpha(z\mid E)}[\log r_\psi(y\mid z)].
\label{eq:label_bound}
\end{equation}

\emph{Rate term (upper bound on $I(E;Z)$).}
Let $\bar q(z)=\E_{p(E)}[q_\alpha(z\mid E)]$ be the aggregated posterior. Using the identity
$\E_{p(E)}\KL(q_\alpha(z\mid E)\Vert p(z))=I(E;Z)+\KL(\bar q(z)\Vert p(z))$
\citep{alemi2017vib} and the non-negativity of $\KL(\bar q\Vert p)$,
\begin{equation}
I(E;Z)\le \E_{p(E)}\KL(q_\alpha(z\mid E)\Vert p(z)).
\label{eq:rate_upper}
\end{equation}

Substituting the lower bound on $I(Y;Z)$ from Eq.~\eqref{eq:label_bound} and the upper bound on $I(E;Z)$ from Eq.~\eqref{eq:rate_upper} into $\mathcal{J}_{\mathrm{EIB}}=I(Y;Z)-\beta I(E;Z)$ yields Eq.~\eqref{eq:eib_bound}.\qed

\subsection{Proof of Proposition~\ref{prop:overconf} (tempering corrects overconfidence under correlated experts)}
\label{app:overconf_proof}
The joint distribution $p(z,\varepsilon_t,\varepsilon_c)$ is Gaussian. Conditioning on $(y_t,y_c)$ and completing the square in $z$ gives the true posterior precision $\tau^*=1+\mathbf{1}^\top\Sigma_\varepsilon^{-1}\mathbf{1}$, where $\Sigma_\varepsilon=\sigma^2\begin{pmatrix}1&\rho\\ \rho&1\end{pmatrix}$ and $\mathbf{1}=(1,1)^\top$. Direct inversion yields $\tau^*=1+2/(\sigma^2(1+\rho))$. For tempered PoE, the diagonal formula of Proposition~\ref{prop:tpoe}, specialized to $d_z=1$, $\sigma_t^2=\sigma_c^2=\sigma^2$, and $\alpha_t=\alpha_c=\alpha$, gives $\tau_\alpha=1+2\alpha/\sigma^2$. Equating $\tau_\alpha=\tau^*$ yields $\alpha^*=1/(1+\rho)$. Monotonicity of $\tau_\alpha$ in $\alpha$ gives the calibration claims.\qed

\subsection{Nuisance leakage consumes bottleneck rate}
\label{app:nuisance}
Suppose evidence decomposes as $E=(S,N)$, where $S$ is label-relevant signal, $N$ is nuisance variation, and $Y\perp N\mid S$. Then
\begin{equation}
I(E;Z)=I(S,N;Z)=I(S;Z)+I(N;Z\mid S).
\end{equation}
The term $I(N;Z\mid S)$ is non-negative and consumes bottleneck rate. Under the evidence-IB objective $I(Y;Z)-\beta I(E;Z)$, nuisance information that reaches $Z$ is penalized but cannot increase the label information once $S$ is known. This does not prove that optimization will always discard nuisance, but it motivates selecting irrelevant components before compression.

\subsection{Diagonal Gaussian KL used for expert disagreement}
\label{app:dskl}
For two diagonal Gaussians $q_A=\N(\mu_A,\diag(\sigma_A^2))$ and $q_B=\N(\mu_B,\diag(\sigma_B^2))$,
\begin{equation}
\KL(q_A\Vert q_B)=\frac{1}{2}\sum_j\left[\log\frac{\sigma_{B,j}^2}{\sigma_{A,j}^2}+\frac{\sigma_{A,j}^2+(\mu_{A,j}-\mu_{B,j})^2}{\sigma_{B,j}^2}-1\right].
\end{equation}
The symmetric expert-conflict diagnostic is
\begin{equation}
D_{\mathrm{SKL}}(q_t,q_c)=\frac{1}{2}\left[\KL(q_t\Vert q_c)+\KL(q_c\Vert q_t)\right].
\end{equation}

\section{Implementation Details}
\label{sec:impl}

The head exposes a small number of interpretable knobs. The bottleneck coefficient $\beta$ controls the evidence information budget and is warmed up over $T_{\mathrm{warm}}$ epochs. The sparsity and consistency coefficients $\eta$ and $\gamma$ control gate mass and stability. The temperatures $\alpha_t,\alpha_c\in[0,1]$ discount correlated evidence axes, and the latent dimension $d_z$ sets the rate ceiling of the decision path. Temporal and channel gates are two-layer MLPs with hidden width 64 and GELU activations. Each Gaussian expert uses a linear--GELU--linear projection to produce $\mu$ and $\log\sigma^2$, with log-variance clamped to $[-10,10]$ for numerical stability. Unless otherwise stated, $d_z=64$.

We evaluate two tempered configurations: a fixed variant with $\alpha_t=\alpha_c=0.5$ and a learnable variant in which $\alpha_t,\alpha_c$ are constrained to $[0,1]$ by sigmoid reparameterization initialized at 0.5. Standard non-tempered PoE is recovered by setting $\alpha_t=\alpha_c=1.0$ and is used as an ablation.

At evaluation time, all stochastic bottlenecks are made deterministic by setting $z=\mu_\alpha$. This makes insertion--deletion curves, posterior reliance, and expert-drop diagnostics reproducible. Expert-drop changes only posterior aggregation weights; it does not mask the EEG input.

\begin{table}[htbp]
\centering
\caption{Default head and optimization hyperparameters used unless otherwise stated.}
\label{tab:hyperparams}
\small
\begin{tabular}{@{}ll@{}}
\toprule
Parameter & Value \\
\midrule
Latent dimension $d_z$ & 64 \\
Bottleneck coefficient $\beta$ & $10^{-4}$ \\
Warmup $T_{\mathrm{warm}}$ & 5 epochs \\
Sparsity coefficient $\eta$ & $10^{-3}$ \\
Consistency coefficient $\gamma$ & $10^{-2}$ \\
Optimizer & AdamW \\
Learning rate & $10^{-3}$ \\
Weight decay & $10^{-3}$ \\
Maximum epochs & 50 \\
Early-stopping patience & 15 epochs \\

\bottomrule
\end{tabular}
\end{table}

\section{Dataset Details}
\label{app:datasets}

Table~\ref{tab:datasets} reports the raw dataset statistics and split conventions used in Sec.~\ref{sec:experiments}. In addition to TUH-derived public benchmarks, we use private resting-state clinical EEG cohorts to test disease-oriented diagnosis under the same frozen-backbone adaptation setting. These in-house labels reflect routine clinical assessment informed by EEG review and chart-level context rather than waveform annotation alone. The depression and cerebrovascular cohorts are binary condition-vs-control datasets built from de-identified 5~s windows; AD staging is a four-way AD/MCI/SCD/HC cohort. Together, the in-house cohorts provide a denser channel axis ($58$--$129$ EEG channels post-preprocessing) than the public benchmarks ($16$--$22$) and therefore a more demanding setting for channel-faithfulness analysis.

\paragraph{Preprocessing.}
We follow the CBraMod~\citep{wang2024cbramod} preprocessing protocol, which is also the convention used by CodeBrain~\citep{ma2025codebrain}: all recordings are uniformly resampled to $200$~Hz before windowing so that the temporal axis is aligned across heterogeneous public and in-house cohorts (the ``$f_s$'' column of Table~\ref{tab:datasets} reports the original acquisition rates). Channel selection is dataset-specific. (i)~The cerebrovascular cohort is acquired with a $140$-channel EGI net; we retain the $129$ EEG-bearing electrodes and drop the $11$ non-EEG sensors (reference / face / EOG / ground leads) before windowing. (ii)~The depression cohort mixes EGI ($129$-channel HydroCel) and BrainVision ($64$-channel) acquisitions; we first align the two systems on their common $10$--$20$-positioned electrodes, then retain the $63$ EEG channels shared by both montages and exclude the BrainVision ECG lead. (iii)~The AD-staging cohort is acquired with EGI on a $58$-electrode grid and used as-is. The TUH public benchmarks (TUAB, TUEV, TUSZ) follow the standard CBraMod bipolar montage and label-window construction unchanged. The ``Ch.'' column in Table~\ref{tab:datasets} therefore lists the maximum acquisition channel count of each cohort; the channel grid actually fed to the model after preprocessing is $129$ for cerebrovascular, $63$ for depression, and matches the table value for the remaining tasks.

All in-house data are private and outside any pretraining corpus. Only de-identified windowed EEG segments and derived record metadata enter the modeling pipeline; institution-identifying ethics and review-board details are not included in this public version. Splits are constructed before windowing so that no subject contributes segments to more than one split. De-identified in-house data may be made available for research use upon reasonable request, subject to approval by the responsible investigators and applicable institutional data-governance constraints. We do not provide formal site or demographic confounding audits for these private cohorts, so in-house results should be read with that caveat.

\begin{table}[htbp]
\centering
\caption{Raw dataset statistics. Splits are reported in number of windows;
the ``Ch.'' column gives the maximum acquisition channel count and ``$f_s$''
the original sampling rate (Hz). All recordings are uniformly resampled to
$200$~Hz before windowing; channel selection (cerebrovascular $\!140\!\to\!129$,
depression $\!64\!\to\!63$ after EGI/BrainVision alignment) is described above.}
\label{tab:datasets}
\small
\setlength{\tabcolsep}{4pt}
\begin{tabular}{@{}llcrlcc@{}}
\toprule
Dataset & Task & $|\mathcal{Y}|$ & Subj. & Train/Val/Test & Ch. & $f_s$ \\
\midrule
TUAB            & Abnormality      & 2 & 2{,}329 & 297k/75k/37k    & 16  & 250/256 \\
TUEV            & Event type       & 6 & 294     & 176k/---/29k    & 16  & 250 \\
TUSZ            & Seizure          & 2 & 675     & 542k/---/92k    & 22  & 200 \\
AD staging      & Cognitive decline & 4 & 108     & 11.7k/2.5k/2.5k & 58  & 250 \\
Depression      & MDD vs.\ control & 2 & 159     & 4.6k/1.0k/0.6k  & 64  & 500/1000 \\
Cerebrovascular & Stroke / lesion  & 2 & 83      & 7.4k/1.5k/1.8k  & 140 & 500/5000 \\
\bottomrule
\end{tabular}
\end{table}

\section{Evaluation Protocol Details}
\label{app:eval_protocol}

This appendix expands the audit protocols summarized in Section~\ref{sec:eval}.

\paragraph{Insertion--deletion gap (extended).}
For each axis $a\in\{t,c\}$, we rank temporal segments or channels by descending gate value. Insertion starts from a fully zero-masked input and progressively restores high-ranked segments or channels; deletion starts from the original input and progressively replaces the same ranked units with zeros. The full model is re-run after each perturbation, while the ranking is fixed from the clean input. Mask fractions are uniformly spaced on $\{0,0.1,\dots,1.0\}$. For each fraction, we record the predicted target-class probability, average over evaluation samples, and integrate to obtain $\mathrm{AUC}^{(a)}_{\mathrm{ins}}$ and $\mathrm{AUC}^{(a)}_{\mathrm{del}}$. The gap is $\mathrm{gap}_a=\mathrm{AUC}^{(a)}_{\mathrm{ins}}-\mathrm{AUC}^{(a)}_{\mathrm{del}}$. To connect selection with posterior integration, we additionally stratify samples by posterior reliance: if high-$\rho_t$ samples show larger temporal gaps than low-$\rho_t$ samples, temporal reliance aligns with temporal gate faithfulness.

\paragraph{Expert necessity (extended).}
Setting both temperatures to zero gives the prior-only noise floor $q_\alpha=p(z)$. We use balanced accuracy as the predictive metric $M$ throughout; $\Delta$ values reported in tables are averaged over three seeds.

\paragraph{Gate causality (extended).}
The override sweep is evaluated on the same held-out test set as the predictive metric. We use 8 sweep values $v\in\{0.0,0.3,0.5,0.6,0.7,0.8,0.9,0.999\}$, with finer resolution near the typical learned-gate regime. The complement of the resulting $M(v)$ curve is reported for CVD and Depression in Fig.~\ref{fig:gate_causality_appendix}.

\paragraph{Symmetric KL between experts (diagonal-Gaussian form).}
For two diagonal-covariance Gaussians $q_t=\mathcal{N}(\mu_t,\diag(\sigma_t^2))$ and $q_c=\mathcal{N}(\mu_c,\diag(\sigma_c^2))$,
\begin{equation*}
\KL(q_t\Vert q_c)=\frac{1}{2}\sum_j\left[\log\frac{\sigma_{c,j}^2}{\sigma_{t,j}^2}+\frac{\sigma_{t,j}^2+(\mu_{t,j}-\mu_{c,j})^2}{\sigma_{c,j}^2}-1\right],
\end{equation*}
and $D_{\mathrm{SKL}}=\tfrac{1}{2}[\KL(q_t\Vert q_c)+\KL(q_c\Vert q_t)]$. We report sample-level $D_{\mathrm{SKL}}$ and its conditional means under correct/incorrect prediction subsets.

\section{Reporting Details}

The evaluation records both summary and per-sample quantities. Summary reports include predictive metrics, calibration, terminal KL, evidence gaps with reliance stratification, expert-drop deltas, posterior reliance means, and expert disagreement. Per-sample reports include labels, predictions, confidence, $\mathrm{gap}_t$, $\mathrm{gap}_c$, $\rho_t$, $\rho_c$, $\rho_0$, $D_{\mathrm{SKL}}(q_t,q_c)$, and posterior variance. These quantities separate accuracy, gate faithfulness, posterior reliance, and temporal--channel conflict, enabling audits of both evidence selection and posterior integration.

\section{Additional Experimental Results}

\subsection{Reliance-stratified gap}
\label{app:reliance_stratified}

\begin{figure}[htbp]
\centering
\includegraphics[width=0.96\linewidth]{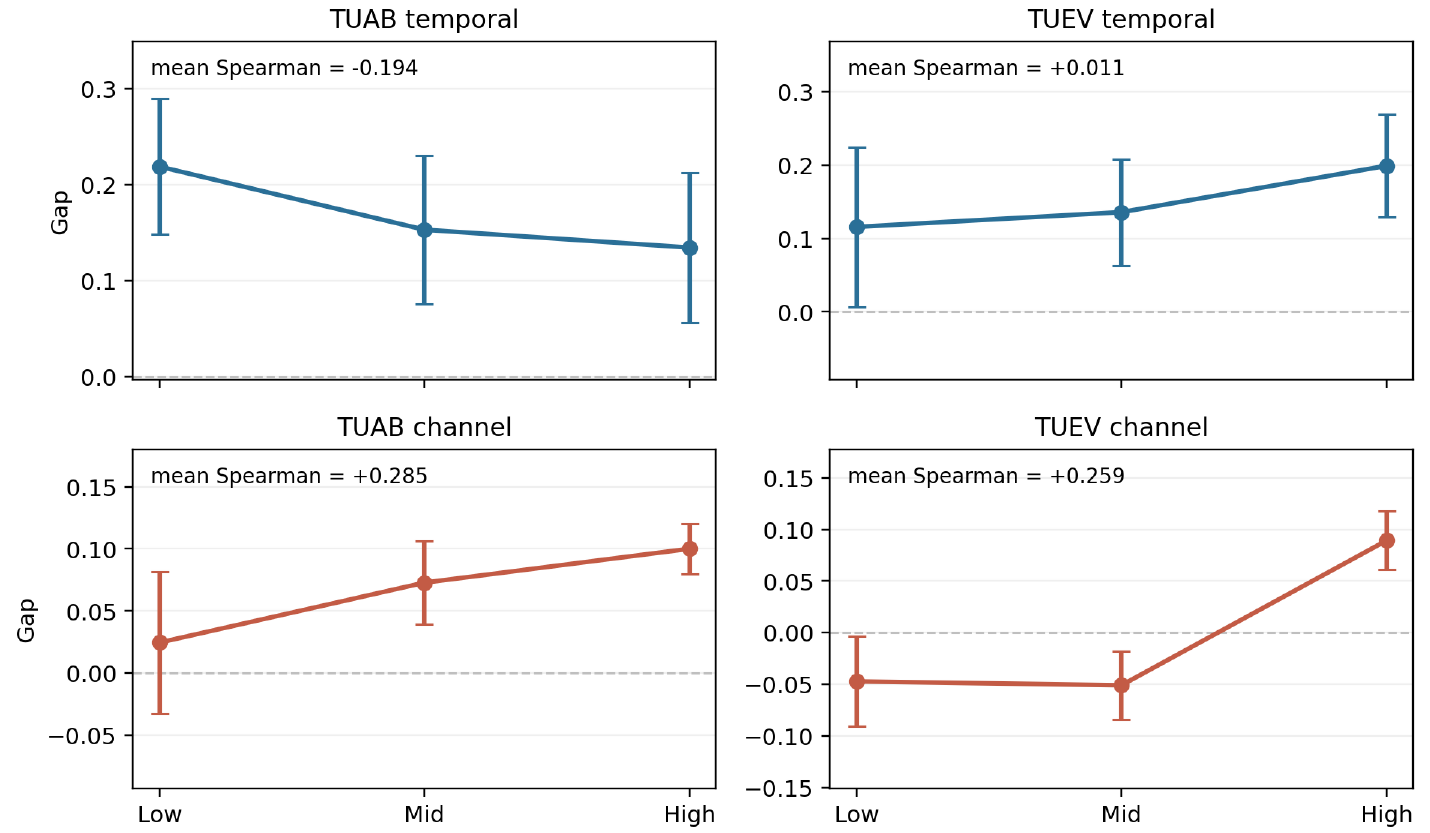}
\caption{Gate-faithfulness gap stratified by posterior-reliance tertile for the learnable-$\alpha$ tPoE-EIB variant on TUAB and TUEV. Each panel reports the mean gap in low/mid/high reliance groups, averaged over three seeds. Higher $\rho_c$ aligns with larger $\mathrm{gap}_c$ on both datasets, whereas temporal alignment is task-dependent.}
\end{figure}

The clearest alignment is on the channel axis: in both TUAB and TUEV, higher $\rho_c$ corresponds to larger $\mathrm{gap}_c$, so samples that receive more posterior precision from the channel expert tend to show more faithful channel perturbation rankings. Temporal alignment is task-dependent: TUEV shows an average increase in $\mathrm{gap}_t$ from low to high $\rho_t$, whereas TUAB does not. On TUAB, the temporal-reliance--gap correlation is weakly negative ($r\approx-0.19$), indicating that on this dataset temporal reliance reflects model confidence rather than perturbation faithfulness.

\subsection{Full calibration table}
\label{app:calibration_full}

\begin{table}[htbp]
\centering
\caption{Full calibration diagnostics across all four heads, public benchmarks. Terminal KL is the final training KL for bottlenecked heads.}
\small
\begin{tabular}{@{}llccc@{}}
\toprule
Dataset & Method & ECE $\downarrow$ & NLL $\downarrow$ & Terminal KL \\
\midrule
TUAB & Linear probe & 0.024 & 0.449 & 0.0 \\
 & Axis-Concat & \textbf{0.018} & \textbf{0.425} & 0.0 \\
 & PoE-EIB ($\alpha=1$) & 0.030 & 0.429 & 48.1 \\
 & tPoE-EIB (learnable $\alpha$) & 0.029 & 0.427 & 43.9 \\
\midrule
TUEV & Linear probe & 0.126 & 0.895 & 0.0 \\
 & Axis-Concat & 0.139 & 1.220 & 0.0 \\
 & PoE-EIB ($\alpha=1$) & 0.138 & 1.114 & 30.0 \\
 & tPoE-EIB (learnable $\alpha$) & \textbf{0.117} & \textbf{0.882} & 28.2 \\
\midrule
TUSZ & Linear probe & 0.051 & 0.299 & 0.0 \\
 & Axis-Concat & 0.026 & 0.272 & 0.0 \\
 & PoE-EIB ($\alpha=1$) & 0.021 & 0.255 & 39.2 \\
 & tPoE-EIB (learnable $\alpha$) & \textbf{0.018} & \textbf{0.227} & 39.2 \\
\bottomrule
\end{tabular}
\end{table}
\label{app:additional_experiments}

\subsection{Cross-backbone generalization (full table)}
\label{app:cbramod}

\begin{table}[htbp]
\centering
\caption{Balanced accuracy (\%) on TUAB with a frozen CBraMod backbone (full four-head sweep). Mean $\pm$ std over 3 seeds. Companion to Sec.~\ref{subsec:main_results}; CBraMod TUEV results (Linear and tPoE-EIB only, 6-class) are reported in Table~\ref{tab:external_baselines}.}
\label{tab:cbramod}
\small
\begin{tabular}{@{}lcccc@{}}
\toprule
Dataset & Linear probe & Axis-Concat & PoE-EIB ($\alpha=1$) & tPoE-EIB (learnable $\alpha$) \\
\midrule
TUAB & $68.2{\pm}0.6$ & $\underline{72.0{\pm}0.2}$ & $71.8{\pm}0.6$ & $\mathbf{72.5{\pm}0.2}$ \\
\bottomrule
\end{tabular}
\end{table}

On TUAB, CBraMod's lower absolute performance compresses the gap among heads, but the relative ordering is preserved: tPoE-EIB ($72.5\%$) $>$ Axis-Concat ($72.0\%$) $>$ PoE-EIB ($71.8\%$) $\gg$ Linear ($68.2\%$). On TUEV (6-class), the two completed CBraMod runs in Table~\ref{tab:external_baselines} show the same Linear $\to$ tPoE-EIB direction ($30.4\!\to\!33.1\%$, $+2.7$) as on CodeBrain ($46.7\!\to\!48.8\%$, $+2.1$), with the absolute gain larger on the weaker CBraMod backbone---consistent with the main-text observation that the evidence interface is most valuable when the underlying representation is weaker.

\subsection{Temperature ablation}
\label{app:temperature}

Table~\ref{tab:temperature} compares fixed $\alpha=0.5$, fixed $\alpha=1.0$, and a learnable $\alpha\in[0,1]$ via sigmoid reparameterization. The learnable variant attains the highest balanced accuracy on all three tasks and the lowest ECE on TUSZ and TUEV; standard PoE ($\alpha=1.0$) gives the lowest ECE on TUAB. Tempering therefore trades calibration for accuracy on event-centric and balanced tasks, and recovers both on the most imbalanced setting (TUSZ and TUEV).

\begin{table}[htbp]
\centering
\caption{Temperature ablation. Each cell reports balanced accuracy (\%) / ECE.}
\label{tab:temperature}
\small
\begin{tabular}{@{}lccc@{}}
\toprule
Variant & TUAB & TUEV & TUSZ \\
\midrule
$\alpha=0.5$ (fixed) & 80.2 / 0.031 & 46.8 / 0.135 & 75.7 / 0.021 \\
$\alpha=1.0$ (standard PoE) & 80.3 / \textbf{0.028} & {47.1} / 0.128 & 75.9 / 0.021 \\
Learnable $\alpha$ & \textbf{80.7} / 0.031 & \textbf{48.8} / \textbf{0.117} & \textbf{76.4} / \textbf{0.018} \\
\bottomrule
\end{tabular}
\end{table}

\subsection{Loss ablation}
\label{app:loss_ablation}

We remove each regularizer in $\mathcal{L}_{\mathrm{total}}$ in turn (Table~\ref{tab:loss_ablation}). The pattern is channel-count-dependent. On TUAB, single-regularizer ablations change BAcc by at most 0.5 pts and leave $\mathrm{gap}_c$ positive but small (\,$+0.051$ to $+0.061$\,), so the regularizers reshape the evidence interface only modestly when the channel count is small. On the 129-channel Depression cohort, the regularizers are essential for channel compression: removing any single regularizer flips $\mathrm{gap}_c$ sharply negative ($+0.104$ with the full loss vs.\ $-0.113$, $-0.158$, $-0.190$ when removing $\mathcal{L}_{\mathrm{sparse}}$, $\mathcal{L}_{\mathrm{cons}}$, $\mathcal{L}_{\mathrm{KL}}$ respectively), with BAcc largely preserved. On TUEV (6-class), single-regularizer ablations cost $2$--$3$ pts of BAcc, and removing $\mathcal{L}_{\mathrm{KL}}$ flips $\mathrm{gap}_c$ from $-0.003$ to $+0.037$. Terminal KL on TUEV is reported on a per-sample basis throughout this ablation, matching the per-example coding-rate interpretation that motivates the bottleneck; since every row of the table uses this same convention, the sign and ranking of within-table differences are invariant to the normalization choice. The contrast between TUAB and Depression suggests channel-axis compression matters more in dense-channel settings, while temporal-axis sparsity/consistency penalties carry more accuracy weight on multi-class event detection.

\begin{table}[htbp]
\centering
\caption{Loss ablation on TUAB, TUEV, and Depression.}
\label{tab:loss_ablation}
\small
\begin{tabular}{@{}llccccc@{}}
\toprule
Dataset & Variant & Bal.\ Acc.\ $\uparrow$ & ECE $\downarrow$ & Terminal KL & $\mathrm{gap}_t$ & $\mathrm{gap}_c$ \\
\midrule
\multirow{4}{*}{TUAB} & Full & 80.7 & $0.029$ & $43.9$ & $+0.161$ & $+0.061$ \\
                     & w/o $\mathcal{L}_{\mathrm{sparse}}$ & 80.2 & $0.028$ & $45.6$ & $+0.153$ & $+0.059$ \\
                     & w/o $\mathcal{L}_{\mathrm{cons}}$ & 80.6 & $0.022$ & $43.9$ & $+0.148$ & $+0.051$ \\
                     & w/o $\mathcal{L}_{\mathrm{KL}}$ (axcat) & 80.6 & $0.030$ & $-$ & $+0.154$ & $+0.056$ \\
\midrule
\multirow{4}{*}{TUEV} & Full & 48.8 & $0.117$ & $106.2$ & $+0.150$ & $-0.003$ \\
                     & w/o $\mathcal{L}_{\mathrm{sparse}}$ & 45.4 & $0.232$ & $109.8$ & $0.131$ & $-0.012$ \\
                     & w/o $\mathcal{L}_{\mathrm{cons}}$ & 46.5 & $0.217$ & $105.8$ & $0.057$ & $-0.114$ \\
                     & w/o $\mathcal{L}_{\mathrm{KL}}$ (axcat) & 46.8 & $0.139$ & $-$ & $0.051$ & $0.037$ \\
\midrule
\multirow{4}{*}{Depression} & Full & 58.8 & $0.152$ & $66.4$ & $0.148$ & $0.104$ \\
                     & w/o $\mathcal{L}_{\mathrm{sparse}}$ & 59.2 & $0.122$ & $68.5$ & $0.151$ & $-0.113$ \\
                     & w/o $\mathcal{L}_{\mathrm{cons}}$ & 60.1 & $0.152$ & $66.3$ & $0.244$ & $-0.158$ \\
                     & w/o $\mathcal{L}_{\mathrm{KL}}$ (axcat) & 59.4 & $0.182$ & $-$ & $0.240$ & $-0.190$ \\
\bottomrule
\end{tabular}
\end{table}

\subsection{Gate shaping diagnostics and nuisance probes}
\label{app:gate_nuisance}

Table~\ref{tab:gate_nuisance} summarizes TUAB diagnostics from a matched bottleneck versus no-bottleneck analysis. These are mechanism checks rather than headline benchmarks: they test whether the bottleneck makes gates more selective, less collapsed, and less subject-identifiable. The pattern is axis-dependent, not a universal anti-collapse story. Temporal support becomes less flat, channel support becomes less concentrated, and subject identity is attenuated but not removed from intermediate representations.

\begin{table}[htbp]
\centering
\caption{Supporting TUAB gate-shaping and nuisance diagnostics from a matched bottleneck versus no-bottleneck analysis. Chance subject-ID accuracy is 0.4\%.}
\label{tab:gate_nuisance}
\footnotesize
\begin{tabular}{@{}lccc@{}}
\toprule
Diagnostic & No bottleneck & Bottlenecked VIB & Interpretation \\
\midrule
Temporal gate range & 0.137 & 0.316 & less flat, more selective temporal support \\
Coverage above 0.5 & 0.002 & 0.207 & high-confidence temporal support emerges \\
Channel Gini & 0.472 & 0.358 & less concentrated channel weighting \\
Effective channels & 10.8 & 12.7 & broader channel support, less collapse \\
Subject-ID from channel gate & 43.1\% & 30.9\% & nuisance attenuated, not removed \\
Subject-ID from latent $\mu$ & --- & 21.1\% & residual nuisance survives compression \\
Subject-ID from logits & 0.9\% & 1.1\% & final decision stays near subject chance \\
\bottomrule
\end{tabular}
\end{table}

A descriptive region-level analysis on the same TUAB diagnostic run shows that the bottleneck redistributes channel support away from lateral/fronto-temporal sites and toward centro-parietal/parieto-occipital regions. We do not interpret this as anatomical localization without topographic validation; it is reported only as evidence that the bottleneck changes where channel mass is allocated.

\subsection{Audit summary across all six tasks}
\label{app:audit_full}

Table~\ref{tab:audit_full} extends Table~\ref{tab:audit} with all audit values. 

\begin{table}[htbp]
\centering
\caption{Audit summary across six tasks. }
\label{tab:audit_full}
\small
\begin{tabular}{@{}lcccccccc@{}}
\toprule
Dataset & $\mathrm{gap}_t$ & $\mathrm{gap}_c$ & $\Delta_t$ & $\Delta_c$ & $\rho_t$ & $\rho_c$ & $\rho_0$ & $D_{\mathrm{SKL}}$ \\
\midrule
TUAB & $0.161$ & $0.061$ & $0.226$ & $0.301$ & $0.14$ & $0.24$ & $0.62$ & $445\mathrm{k}$ \\
TUEV & $0.150$ & $-0.003$ & $0.145$ & $0.224$ & $0.18$ & $0.23$ & $0.59$ & $68\mathrm{k}$ \\
TUSZ & $0.098$ & $-0.010$ & $0.130$ & $0.269$ & $0.18$ & $0.30$ & $0.52$ & $42\mathrm{k}$ \\
AD staging & $0.340$ & $-0.172$ & $0.028$ & $0.026$ & $0.43$ & $0.55$ & $0.03$ & $41k$ \\
Depression & $0.148$ & $0.104$ & $0.169$ & $0.153$ & $0.37$ & $0.39$ & $0.24$ & $2k$ \\
Cerebrovascular & $0.196$ & $-0.04$ & $0.037$ & $0.018$ & $0.34$ & $0.38$ & $0.28$ & $2k$ \\
\bottomrule
\end{tabular}
\end{table}

\subsection{Distinguishing axis regimes via gate-causality}
\label{app:gate_causality_regime}

\begin{figure}[htbp]
\centering
\includegraphics[width=0.45\linewidth]{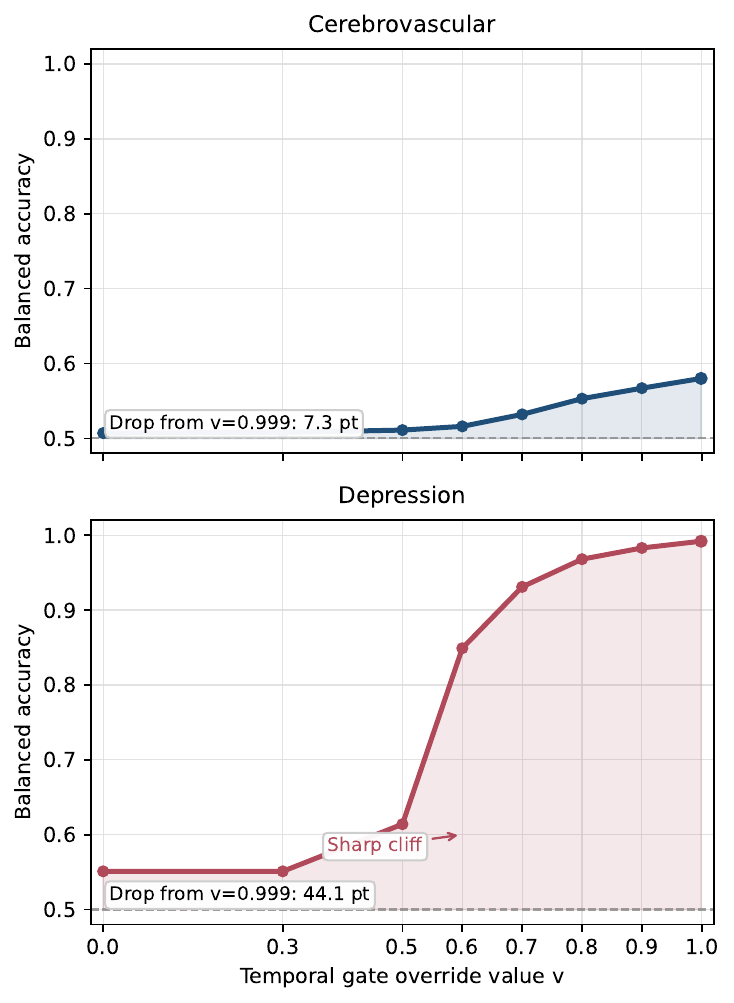}
\caption{Gate-causality sweep (CVD vs.\ Depression). Balanced accuracy under temporal gate override $v$: CVD has a flat response; Depression shows a sharp cliff between $v=0.5$ and $v=0.6$.}
\label{fig:gate_causality_appendix}
\end{figure}

On the CVD dataset, the temporal gate converges to a near-constant high value (mean $0.998$, std $\sim 8{\times}10^{-5}$) while the channel gate retains sample-dependent variation; insertion--deletion is hard to interpret because rankings are nearly tied. The gate-causality sweep (Fig.~\ref{fig:gate_causality_appendix}) resolves the ambiguity: CVD shows a flat response (reducing $v$ from $0.999$ to $0$ costs only $7.3$ pts, no threshold), whereas Depression shows a sharp cliff between $v=0.5$ and $v=0.6$ ($44.1$ pts drop). CVD temporal evidence is therefore weakly causal at the gate stage; Depression requires near-open temporal gating. A nearly constant gate can be irrelevant or uniformly necessary; the sweep separates these cases.

\subsection{Per-task region-level gate distributions}
\label{app:clinical_alignment_figs}

This appendix expands Fig.~\ref{fig:clinical_alignment_main} with the full set of region-level visualizations supporting Table~\ref{tab:clinical_alignment}. Unless stated otherwise, plots are produced from a single seed ($s{=}42$) and the held-out test split. The evaluation pipeline groups channels into canonical 10--20 region sets (frontal: AF/Fp/F; central: FC/C/CP; temporal: T/FT/TP; parietal: P; posterior/occipital: PO/O); for the bipolar TUAB and TUSZ montages the same grouping is applied to derivation pairs by majority vote.

\begin{figure}[htbp]
\centering
\begin{minipage}[b]{0.32\linewidth}\centering\includegraphics[width=\linewidth]{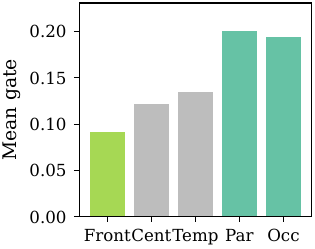}\subcaption{TUAB}\end{minipage}\hfill
\begin{minipage}[b]{0.32\linewidth}\centering\includegraphics[width=\linewidth]{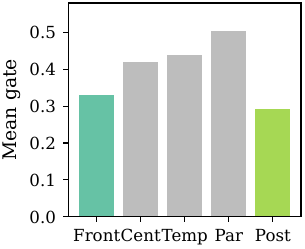}\subcaption{Depression}\end{minipage}\hfill
\begin{minipage}[b]{0.32\linewidth}\centering\includegraphics[width=\linewidth]{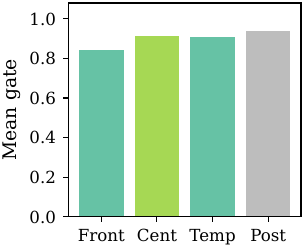}\subcaption{Cerebrovascular}\end{minipage}\\[2pt]
\begin{minipage}[b]{0.32\linewidth}\centering\includegraphics[width=\linewidth]{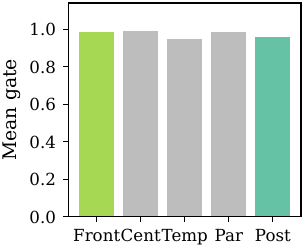}\subcaption{AD staging}\end{minipage}\hfill
\begin{minipage}[b]{0.32\linewidth}\centering\includegraphics[width=\linewidth]{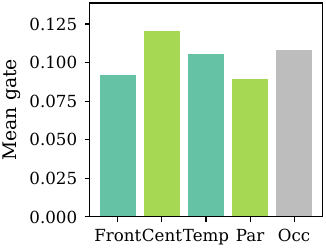}\subcaption{TUSZ}\end{minipage}\hfill
\begin{minipage}[b]{0.32\linewidth}\centering\includegraphics[width=\linewidth]{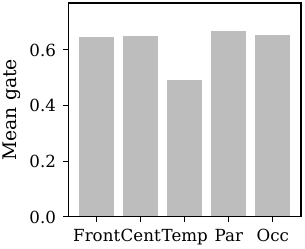}\subcaption{TUEV}\end{minipage}
\caption{Region-averaged mean channel gate per task (single seed $s{=}42$). 
Highlighted bars (teal / lime-green) mark hypothesized target and control 
regions per Table~\ref{tab:clinical_alignment}; gray bars are non-hypothesized 
regions. TUEV uses a per-event-class hypothesis (no single directional 
target), so all bars are gray. The CVD panel uses a 4-region grouping 
without Par/Occ. Note the saturated near-1.0 range on AD, indicating a 
non-discriminative selector at the 4-way granularity, and the 
central/occipital pattern on TUSZ that contradicts the focal-onset prior.}
\label{fig:region_gate_means_grid}
\end{figure}

\paragraph{Per-task region-averaged gate.}
Fig.~\ref{fig:region_gate_means_grid} shows the mean channel gate aggregated 
by region for each task. Highlighted bars mark hypothesized target / control 
regions per Table~\ref{tab:clinical_alignment}; gray bars are non-hypothesized 
regions. TUAB shows a clean parietal/occipital peak with the lowest weight 
on frontal channels; TUSZ shows a central/occipital-dominant profile that 
contradicts the focal-onset (temporal/frontal) prior, consistent with the 
rejected hypothesis in Table~\ref{tab:clinical_alignment}; TUEV shows a 
flat profile across central/frontal/parietal/occipital with a temporal 
dip, consistent with mixed-event heterogeneity; CVD and Depression both 
show a posterior peak in the per-region average; AD shows a saturated 
profile near 1.0 with no resolvable contrast.

\begin{figure}[htbp]
\centering
\begin{minipage}[b]{0.32\linewidth}\centering\includegraphics[width=\linewidth]{tuab_correct_by_class.pdf}\subcaption{TUAB}\end{minipage}\hfill
\begin{minipage}[b]{0.32\linewidth}\centering\includegraphics[width=\linewidth]{dep_correct_by_class.pdf}\subcaption{Depression}\end{minipage}\hfill
\begin{minipage}[b]{0.32\linewidth}\centering\includegraphics[width=\linewidth]{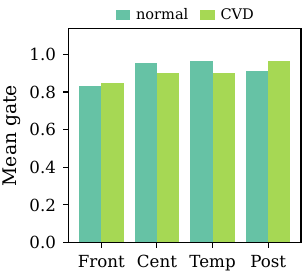}\subcaption{Cerebrovascular}\end{minipage}\\[2pt]
\begin{minipage}[b]{0.32\linewidth}\centering\includegraphics[width=\linewidth]{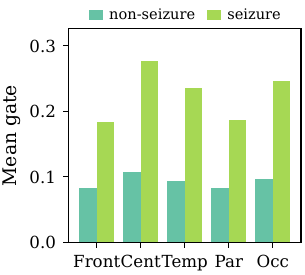}\subcaption{TUSZ}\end{minipage}\hfill
\begin{minipage}[b]{0.32\linewidth}\centering\includegraphics[width=\linewidth]{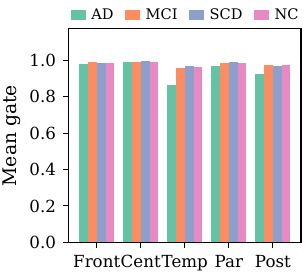}\subcaption{AD staging}\end{minipage}\hfill
\begin{minipage}[b]{0.32\linewidth}\centering\includegraphics[width=\linewidth]{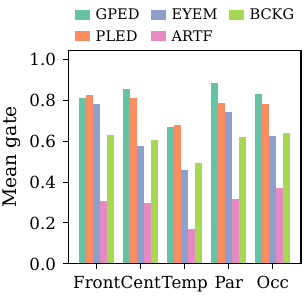}\subcaption{TUEV (per event)}\end{minipage}
\caption{Region-averaged channel gate on correctly classified samples, split by ground-truth class. (a) TUAB: abnormal-class gate $\approx 2{\times}$ normal-class at every region with the largest absolute gap at occipital/parietal. (b) Depression: parietal carries the largest class separation ($0.62$ vs.\ $0.34$). (c) CVD: disease class shifts mass posteriorly (posterior $0.96$ vs.\ $0.91$ on normal) and reduces central/temporal recruitment. (d) TUSZ: seizure class recruits gate $\approx 2$--$3{\times}$ broader than non-seizure, with central/occipital dominance rather than the focal temporal/frontal predicted. (e) AD staging: AD/MCI/SCD/NC profiles are near-identical due to gate saturation. (f) TUEV: EYEM peaks at frontal (eye-movement origin); GPED/PLED at central; ARTF lowest overall.}
\label{fig:correct_by_class_grid}
\end{figure}

\paragraph{Correct-only, class-conditioned profiles (used in Fig.~\ref{fig:clinical_alignment_main}).}
Fig.~\ref{fig:correct_by_class_grid} restricts the analysis to samples the model classified correctly and then splits by ground-truth class. This view removes both the class-imbalance confound (the dominant class also contributes most correct predictions) and the prediction-error confound (selector behavior on misclassified samples is hard to interpret). The resulting differences are direct evidence that the selector encodes class-specific anatomical information.

\begin{figure}[htbp]
\centering
\begin{minipage}[b]{0.32\linewidth}\centering\includegraphics[width=\linewidth]{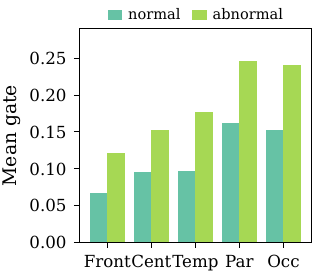}\subcaption{TUAB}\end{minipage}\hfill
\begin{minipage}[b]{0.32\linewidth}\centering\includegraphics[width=\linewidth]{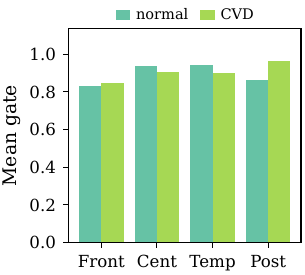}\subcaption{Cerebrovascular}\end{minipage}\hfill
\begin{minipage}[b]{0.32\linewidth}\centering\includegraphics[width=\linewidth]{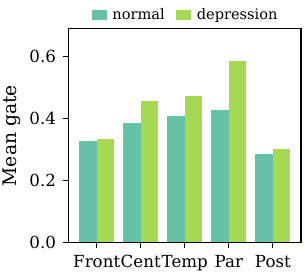}\subcaption{Depression}\end{minipage}\\[2pt]
\begin{minipage}[b]{0.32\linewidth}\centering\includegraphics[width=\linewidth]{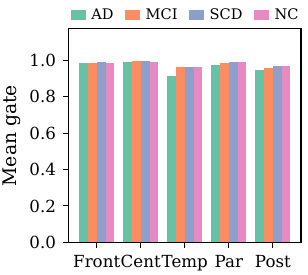}\subcaption{AD staging (4 classes)}\end{minipage}\hfill
\begin{minipage}[b]{0.32\linewidth}\centering\includegraphics[width=\linewidth]{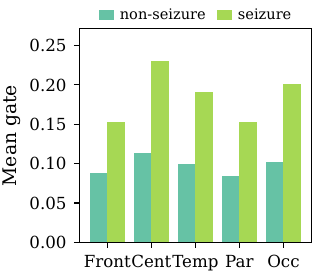}\subcaption{TUSZ}\end{minipage}\hfill
\begin{minipage}[b]{0.32\linewidth}\centering\includegraphics[width=\linewidth]{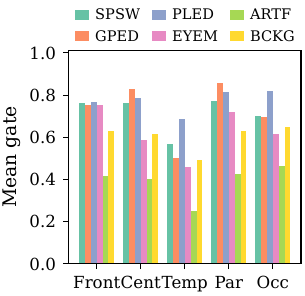}\subcaption{TUEV (6 events)}\end{minipage}
\caption{Per-class regional gate profiles on all test samples (no correct-only filter). Each bar gives the mean channel gate within the region, conditioned on ground-truth class.}
\label{fig:class_profiles_grid}
\end{figure}

\paragraph{Per-class regional profiles (all samples).}
Fig.~\ref{fig:class_profiles_grid} stratifies the regional profile by ground-truth class on \emph{all} test samples (i.e., without the correct-only filter used in Fig.~\ref{fig:correct_by_class_grid}). TUAB shows the abnormal class consistently above the normal class at every region, with the largest gap at parietal/occipital ($0.24$--$0.25$ vs.\ $0.15$--$0.16$), mirroring the correct-only view in Fig.~\ref{fig:correct_by_class_grid}a; CVD shows a clear posterior shift in the disease class ($0.96$) relative to normal controls ($0.86$) and a complementary central/temporal reduction; Depression shows broader elevation at central, parietal, and temporal sites for the depressive class, with parietal carrying the largest class separation ($0.58$ vs.\ $0.43$); AD class profiles are visibly compressed (all five regions within $\sim 0.05$ of $1.0$ across AD/MCI/SCD/NC); TUSZ separates seizure from non-seizure markedly at every region (seizure $\approx 0.18$--$0.28$ vs.\ non-seizure $\approx 0.08$--$0.11$); TUEV separates GPED/PLED (broad, high) from ARTF (broad, low) and EYEM (frontal-peaked).

\begin{figure}[htbp]
\centering
\begin{minipage}[b]{0.32\linewidth}\centering\includegraphics[width=\linewidth]{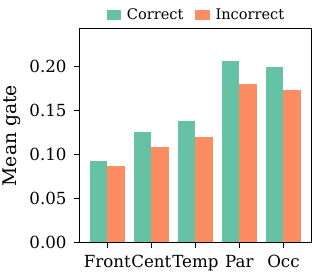}\subcaption{TUAB}\end{minipage}\hfill
\begin{minipage}[b]{0.32\linewidth}\centering\includegraphics[width=\linewidth]{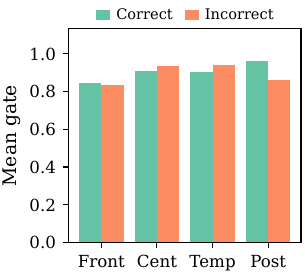}\subcaption{Cerebrovascular}\end{minipage}\hfill
\begin{minipage}[b]{0.32\linewidth}\centering\includegraphics[width=\linewidth]{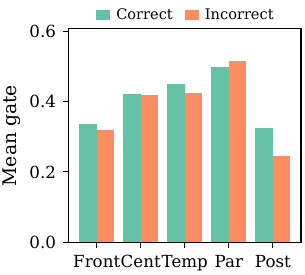}\subcaption{Depression}\end{minipage}\\[2pt]
\begin{minipage}[b]{0.32\linewidth}\centering\includegraphics[width=\linewidth]{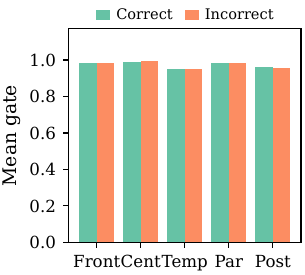}\subcaption{AD staging}\end{minipage}\hfill
\begin{minipage}[b]{0.32\linewidth}\centering\includegraphics[width=\linewidth]{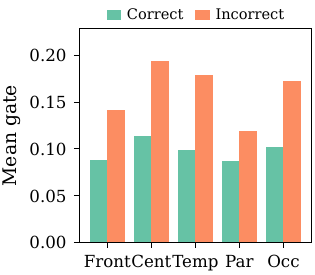}\subcaption{TUSZ}\end{minipage}\hfill
\begin{minipage}[b]{0.32\linewidth}\centering\includegraphics[width=\linewidth]{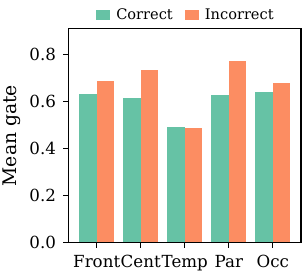}\subcaption{TUEV}\end{minipage}
\caption{Region-averaged channel gate, split by prediction outcome. AD shows no separation under the saturated 4-way gate; TUSZ shows the opposite pattern from TUAB/CVD/Depression, with incorrect samples carrying higher gate.}
\label{fig:correct_incorrect_grid}
\end{figure}

\paragraph{Correct vs.\ incorrect breakdown.}
Fig.~\ref{fig:correct_incorrect_grid} compares the regional gate profile of correctly classified samples against incorrectly classified ones, irrespective of class. On TUAB, CVD, and Depression, the correctly predicted subset shows higher gate at the clinically expected target region. On TUSZ the pattern reverses: incorrectly classified samples have markedly higher gate everywhere, suggesting the selector over-recruits when the input is ambiguous. On the imbalanced CVD and Depression splits the contrast partly reflects the fact that the dominant disease class is also the better-classified one, so this view should be read together with Fig.~\ref{fig:correct_by_class_grid}. AD shows nearly identical profiles between correct and incorrect, mirroring the saturated 4-way gate even though balanced accuracy is best-in-class.

\begin{figure}[htbp]
\centering
\begin{minipage}[b]{0.32\linewidth}\centering\includegraphics[height=2.8cm]{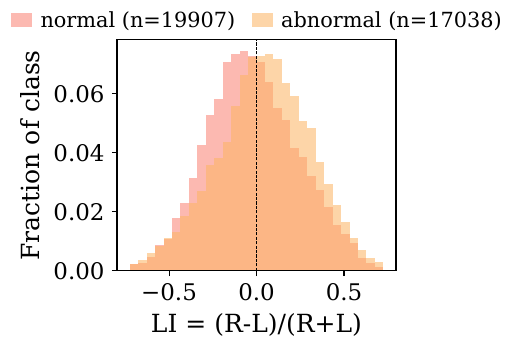}\subcaption{TUAB}\end{minipage}\hfill
\begin{minipage}[b]{0.32\linewidth}\centering\includegraphics[height=2.8cm]{cvd_lateralization.pdf}\subcaption{Cerebrovascular}\end{minipage}\hfill
\begin{minipage}[b]{0.32\linewidth}\centering\includegraphics[height=2.8cm]{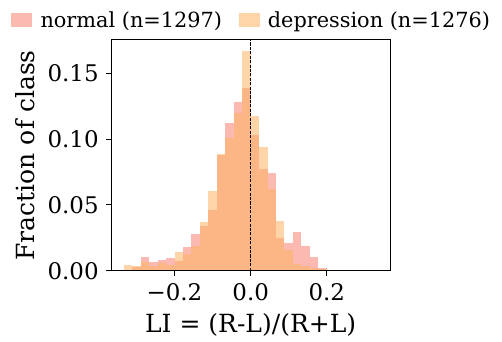}\subcaption{Depression}\end{minipage}\\[4pt]
\begin{minipage}[b]{0.32\linewidth}\centering\includegraphics[height=2.8cm]{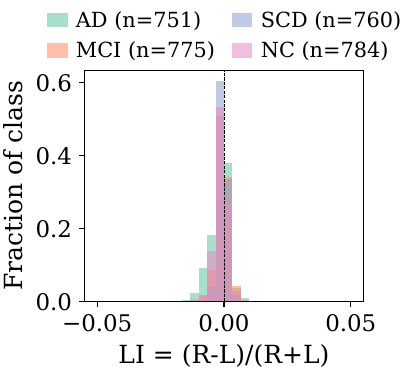}\subcaption{AD staging}\end{minipage}\hfill
\begin{minipage}[b]{0.32\linewidth}\centering\includegraphics[height=2.8cm]{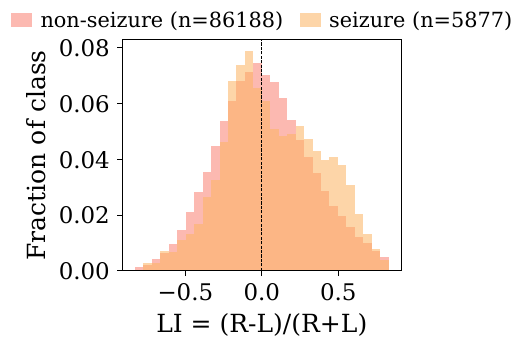}\subcaption{TUSZ}\end{minipage}\hfill
\begin{minipage}[b]{0.32\linewidth}\centering\includegraphics[height=2.8cm]{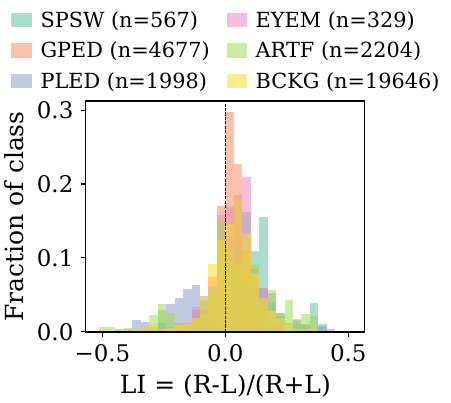}\subcaption{TUEV}\end{minipage}
\caption{Per-sample lateralization index $\mathrm{LI}=(L-R)/(L+R)$. TUAB: normal and abnormal distributions overlap with a small class-conditional shift, consistent with PDR being bilateral. CVD: clear broadening of the distribution for the disease class. Depression: modest left-bias consistent with weak FAA. AD: concentrated near zero on a narrow $\pm 0.015$ axis---no recovered lateralization. TUSZ/TUEV: shown for completeness.}
\label{fig:lateralization_grid}
\end{figure}

\paragraph{Lateralization index.}
Fig.~\ref{fig:lateralization_grid} shows the per-sample lateralization index $\mathrm{LI}=(L-R)/(L+R)$ histogram for tasks with a defined class structure. On TUAB, normal and abnormal class distributions are broad and largely overlapping, with only a small class shift (normal peaks at $\mathrm{LI}\!\approx\!-0.05$, abnormal at $\mathrm{LI}\!\approx\!+0.05$), consistent with PDR being a bilateral posterior phenomenon rather than a lateralized one. The CVD class shows a substantially broader distribution than normal controls (Mann--Whitney $U$ test on $|\mathrm{LI}|$: $p<0.001$, $n_{\text{CVD}}{=}1{,}384$, $n_{\text{normal}}{=}413$), consistent with focal lesion-side slowing rather than a fixed hemispheric bias. Depression shows a modest left-bias relative to normal, weakly consistent with the FAA prediction (more left-frontal alpha $\rightarrow$ negative $\mathrm{LI}$). On AD the histogram is concentrated near zero on a $\pm 0.015$ axis, indicating that the selector does not recover a lateralized pattern even within the AD class. TUSZ and TUEV are included for completeness; for TUSZ the seizure class shows a wider tail than non-seizure but no consistent hemispheric bias.

\subsection{AD vs.\ NC binary reformulation}
\label{app:ad_nc_binary}

The four-way AD/MCI/SCD/NC staging task in Sec.~\ref{subsec:clinical_alignment} is the hardest in-house setting: although tPoE-EIB attains the highest balanced accuracy among heads ($38.5{\pm}4.1\%$, chance $25\%$), the per-region channel gate is saturated near $1.0$ and the alignment audit is rejected. To test whether this reflects a fundamental inability of the bottleneck to capture AD-relevant evidence in this cohort or simply task granularity, we re-train tPoE-EIB on a binary AD vs.\ NC reformulation of the same cohort (same 58-channel grid, same frozen CodeBrain backbone, same default hyperparameters; subject-disjoint splits restricted to AD and NC subjects, with $n_{\mathrm{AD}}{=}751$ and $n_{\mathrm{NC}}{=}784$ test windows). We then probe the learned channel-gate space with a logistic regression (LR) on the 58-dimensional channel gate $g_c(x)$ and characterize differential gate behaviour across three seeds.

\paragraph{Linear separability of the channel-gate space.}
Table~\ref{tab:ad_nc_lr} reports balanced accuracy of a logistic regression trained on $g_c(x)$ produced by the tPoE-EIB head, separately for each training seed. The LR reaches $88.8$--$97.9\%$ balanced accuracy, indicating that the channel-gate space is approximately linearly separable for AD vs.\ NC even though the bottlenecked classifier rate-limits the information that can flow from the gates to the prediction. This complements rather than replaces the four-way staging result: at finer label granularity the bottlenecked classifier loses discriminative resolution; at the binary level the gate space already encodes class-discriminative spatial information.

\begin{table}[htbp]
\centering
\caption{Logistic regression on the 58-dimensional channel gate $g_c(x)$ produced by tPoE-EIB on the binary AD vs.\ NC reformulation, evaluated per training seed. Top channels are listed by absolute LR weight; sign in parentheses indicates the direction of the AD vs.\ NC contrast (negative = NC favoured by larger gate value).}
\label{tab:ad_nc_lr}
\small
\begin{tabular}{@{}lcl@{}}
\toprule
Seed & LR BAcc & Top weighted channels \\
\midrule
s42   & $0.948$ & POz$(-)$, CP2$(-)$, PO4$(+)$, P8$(+)$, TP8$(-)$ \\
s2025 & $0.979$ & T8$(-)$, O2$(-)$, TP8$(-)$, AF3$(-)$, F3$(-)$ \\
s3407 & $0.888$ & CP2$(-)$, O1$(+)$, TP7$(+)$, AF4$(-)$, F3$(-)$ \\
\bottomrule
\end{tabular}
\end{table}

\paragraph{Differential channel-gate behaviour.}
Mann--Whitney $U$ tests per channel identify channels whose mean gate differs between AD and NC. Across seeds, NC consistently shows higher gate activation than AD on most differentially gated channels: 3-seed mean differences are negative for the top 15 differentiated channels, with absolute gaps of $0.13$--$0.23$. The differentially gated channels span parietal (P2, P5, P6, P7), occipital (O2), temporal--parietal (TP7, TP8), central (C1, C2, C6), and frontal (F3, AF3, AF4, F8, FC5) sites rather than concentrating at a single locus. The pattern is therefore inconsistent with a clean ``posterior $>$ frontal'' rule but consistent with AD as a diffuse rather than focal process.

\paragraph{Fronto-parietal correlation reversal.}
The most pronounced AD$-$NC differences in pairwise correlations of channel-gate values occur between frontal and parietal channels. Across seeds, the F3--P3, F8--P3, F4--P3, FC5--P3, F8--P6, Fz--P6, FC5--P6, and F7--P7 pairs all show gate-value correlations that flip sign from positive in NC to negative in AD, with $\Delta\mathrm{corr}=\mathrm{corr}_{\mathrm{AD}}-\mathrm{corr}_{\mathrm{NC}}\in[-1.13,-0.84]$ on average. This fronto-parietal decoupling at the gate level is not built into the architecture; it emerges from the data and is broadly consistent with diffuse fronto-parietal disruption reported in the AD EEG literature \citep{jeong2004eegad,lizio2011eegad}.

\paragraph{Temporal-gate behaviour.}
Temporal-gate statistics are seed-dependent. In seed s42 the AD class shows a higher temporal-gate mean ($0.877$ vs.\ $0.826$, $p < 10^{-22}$); in seed s2025 the direction reverses ($0.970$ vs.\ $0.976$, $p \approx 10^{-3}$); in seed s3407 the temporal gate saturates near $1.0$ in both classes with no significant difference. Temporal-gate behaviour therefore does not provide a stable AD vs.\ NC marker in this cohort, in contrast to the channel-gate pattern.

\paragraph{Implication.}
The four-way staging failure should be read as task-granularity-driven rather than as evidence that the bottleneck cannot capture AD-relevant features in this cohort. On the binary reformulation, the gate space encodes AD vs.\ NC in a distributed fronto-parietal--occipital pattern with measurable correlation reorganization between frontal and parietal sites. The recovered pattern does not align with the pre-registered ``posterior $>$ frontal'' anchor, so the Q4 alignment verdict for AD remains negative, but the gate is not non-discriminative at the binary level.

\section{Per-sample TUEV gate overlays}
\label{app:tuev_overlays}

\begin{figure}[htbp]
\centering
\begin{minipage}[b]{0.32\linewidth}\centering\includegraphics[width=\linewidth]{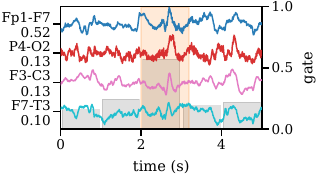}\end{minipage}\hfill
\begin{minipage}[b]{0.32\linewidth}\centering\includegraphics[width=\linewidth]{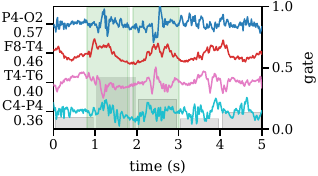}\end{minipage}\hfill
\begin{minipage}[b]{0.32\linewidth}\centering\includegraphics[width=\linewidth]{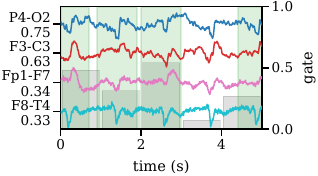}\end{minipage}\\[2pt]
\begin{minipage}[b]{0.32\linewidth}\centering\includegraphics[width=\linewidth]{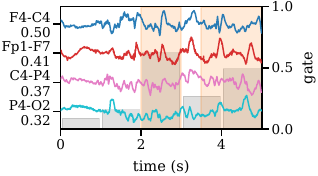}\end{minipage}\hfill
\begin{minipage}[b]{0.32\linewidth}\centering\includegraphics[width=\linewidth]{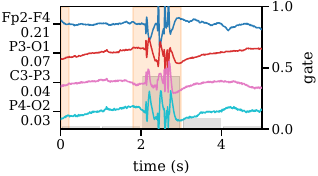}\end{minipage}\hfill
\begin{minipage}[b]{0.32\linewidth}\centering\includegraphics[width=\linewidth]{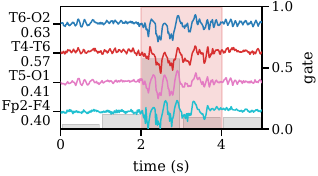}\end{minipage}\\[2pt]
\begin{minipage}[b]{0.32\linewidth}\centering\includegraphics[width=\linewidth]{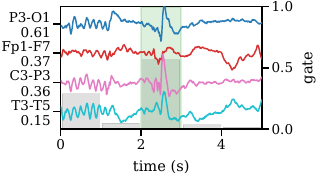}\end{minipage}\hfill
\begin{minipage}[b]{0.32\linewidth}\centering\includegraphics[width=\linewidth]{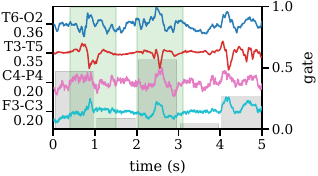}\end{minipage}\hfill
\begin{minipage}[b]{0.32\linewidth}\centering\includegraphics[width=\linewidth]{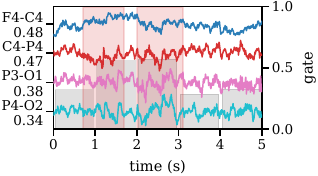}\end{minipage}
\caption{Per-sample TUEV gate overlays on nine correctly-classified pathological events (seed s42). Each panel shows the four highest channel-gate channels with gate weights in parentheses; gray bars mark the temporal-gate selection and the orange band marks the ground-truth event annotation. The selected support concentrates on the event window across diverse event classes and montages.}
\label{fig:tuev_overlays}
\end{figure}

Fig.~\ref{fig:tuev_overlays} shows nine correctly-classified TUEV samples with the trained gates overlaid on the raw signal. Each panel plots the four channels with the highest channel-gate value, annotated with their channel-gate weight; gray bars indicate the temporal-gate selection and the orange shaded region marks the ground-truth event window. The panels illustrate how the temporal gate concentrates around the event annotation while the channel gate selects axes that are visually consistent with the event class (e.g., frontal channels for ocular EYEM events, central/parietal sites for periodic-discharge classes).

\end{document}